\documentclass{aa}  

\usepackage{graphicx}
\usepackage{txfonts}
\usepackage{xcolor}
\usepackage{comment}
\usepackage{footnote}
\makesavenoteenv{tabular}
\makesavenoteenv{table}
\usepackage{soul}
\usepackage{threeparttable}
\usepackage{rotating}
\usepackage{booktabs}
\usepackage{caption}
\usepackage[rightcaption]{sidecap}

\newcommand{\Ha}{H$\alpha$}
\newcommand{\thetaonec}{$\theta^1$ Ori C }
\newcommand{\nodata}{...}

\begin{document} 

% Photoevaporated Protoplanetary Disks in the Orion Nebula Cluster as Seen with MUSE NFM
   \title{{Kaleidoscope of irradiated disks: MUSE observations of proplyds in the Orion Nebula Cluster}
    \thanks{Based on observations collected at the European Southern Observatory under ESO programme 0104.C-0963(A), 106.218X.001, and 110.259E.001.}}
   \subtitle{I. Sample presentation and ionization front sizes}

   \author{M.-L. Aru\inst{\ref{instESO}}
          \and
          K. Mauc\'o\inst{\ref{instESO}}
          \and 
          C. F. Manara\inst{\ref{instESO}}
          \and
          T. J. Haworth\inst{\ref{instQMUL}}
          \and
          S. Facchini\inst{\ref{uniMI}}
          \and
          A. F. McLeod\inst{\ref{Durham},\ref{Durham2}}
          \and
          A. Miotello\inst{\ref{instESO}}
          \and \\
          M. G. Petr-Gotzens\inst{\ref{instESO}}
          \and
          M. Robberto\inst{\ref{STScI},\ref{JHU}}
          \and
          G. P. Rosotti\inst{\ref{uniMI}}
          \and
          S. Vicente\inst{\ref{instPO}}
          \and
          A. Winter\inst{\ref{instOCA},\ref{instIPAG}}
          \and
          M. Ansdell\inst{\ref{NASA}} 
          }

   \institute{European Southern Observatory, Karl-Schwarzschild-Strasse 2, 85748 Garching bei München, Germany\label{instESO}\\
              \email{mariliis.aru@eso.org}
          \and
          Astronomy Unit, School of Physics and Astronomy, Queen Mary University of London, London E1 4NS, UK\label{instQMUL}
          \and
          Dipartimento di Fisica, Università degli Studi di Milano, Via Celoria 16, 20133 Milano, Italy\label{uniMI}
          \and
          Centre for Extragalactic Astronomy, Department of Physics, Durham University, South Road, Durham DH1 3LE, UK\label{Durham}
          \and
          Department of Physics, Institute for Computational Cosmology, University of Durham, South Road, Durham DH1 3LE, UK\label{Durham2}
          \and
          Space Telescope Science Institute, 3700 San Martin Dr, Baltimore, MD 21218, USA\label{STScI}
          \and
          Johns Hopkins University, 3400 N. Charles Street, Baltimore, MD 21218, USA\label{JHU}
          \and
          Instituto de Astrofísica e Ciências do Espaco, Universidade de Lisboa, OAL, Tapada da Ajuda, P-1349-018 Lisboa, Portugal\label{instPO}
          \and
          Universit\'e Cote d'Azur, Observatoire de la Cote d'Azur, CNRS, Laboratoire Lagrange, F-06300 Nice, France\label{instOCA}
          \and
          Universi\'te Grenoble Alpes, CNRS, IPAG, F-38000 Grenoble, France\label{instIPAG}
          \and
          NASA Headquarters, 300 E Street SW, Washington, DC 20546, USA\label{NASA}
         }
         %     }

   \date{Received 18 December 2023. Accepted 2 April 2024.}

% \abstract{}{}{}{}{} 
% 5 {} token are mandatory
 
  \abstract
{In the Orion Nebula Cluster (ONC), protoplanetary disks exhibit ionized gas clouds in the form of a striking teardrop shape as massive stars irradiate the disk material.
We present the first spatially and spectrally resolved observations of 12 such objects, known as proplyds, using integral field spectroscopy observations performed with the Multi-Unit Spectroscopic Explorer (MUSE) instrument in Narrow Field Mode (NFM) on the Very Large Telescope (VLT).
We present the morphology of the proplyds in seven emission lines and measure the radius of the ionization front (I-front) of the targets in four tracers, covering transitions of different ionization states for the same element. We also derive stellar masses for the targets.

The measurements follow a consistent trend of increasing I-front radius for a decreasing strength of the far-UV radiation as expected from photoevaporation models.
By analyzing the ratios of the I-front radii as measured in the emission lines of \Ha, [O\,\textsc{i}] 6300\AA, [O\,\textsc{ii}] 7330\AA, and [O\,\textsc{iii}] 5007\AA, we observe the ionization stratification, that is, the most ionized part of the flow being the furthest from the disk (and closest to the UV source). The ratios of ionization front radii scale in the same way for all proplyds in our sample regardless of the incident radiation. We show that the stratification can help constrain the densities near the I-front by using a 1D photoionization model.

We derive the upper limits of photoevaporative mass-loss rates ($\dot{M}_{\rm loss}$) by assuming ionization equilibrium, and estimate values in the range 1.07--94.5 $\times$ 10$^{-7}$ M$_\odot\,{\rm yr}^{-1}$, with $\dot{M}_{\rm loss}$ values decreasing towards lower impinging radiation. We do not find a correlation between the mass-loss rate and stellar mass. The highest mass-loss rate is for the giant proplyd 244-440. 
These values of $\dot{M}_{\rm loss}$, combined with recent estimates of the disk mass with ALMA, confirm previous estimates of the short lifetime of these proplyds.

This work demonstrates the potential of this MUSE dataset and offers a new set of observables to be used to test current and future models of external photoevaporation.

}
% {}
  % % aims heading (mandatory)
  %  {TBW}
  % % methods heading (mandatory)
  %  {TBW}
  % % results heading (mandatory)
  %  {TBW}
  % % conclusions heading (optional), leave it empty if necessary 
  %  {}

   \keywords{ISM: individual: Orion Nebula -- stars: pre-main-sequence -- protoplanetary disks}

   \maketitle
%
%-------------------------------------------------------------------

\section{Introduction}
Protoplanetary disks, composed of gas and dust, emerge as a consequence of the star formation process, and provide the birthplaces of planetary systems. The evolutionary pathways of protoplanetary disks and their ability to form planets are expected to differ depending on the surrounding environment, with disks undergoing rapid changes in the presence of massive stars (for a recent review see \citealt{Parker2020TheSystems.} and \citealt{Reiter2022DynamicsEnvironments}). In massive clusters near OB-type stars, UV radiation can externally photoevaporate disks and severely diminish their size, mass, and survival timescale \citep[e.g.,][]{WinterHaworth2022}.

The effect of external photoevaporation can be studied in the act in the Orion Nebula Cluster (ONC), where the primary UV source is the O6V star $\theta^1$ Ori C \citep{ODell2017WhichNebula}.
These systems of ionized protoplanetary disks, known as "proplyds", were first imaged with the Hubble Space Telescope (HST), exhibiting ionization fronts, disk silhouettes, and comet-like tails pointing away from the UV source \citep{OdellWen1993,OdellWen1994,McCaughrean1996DirectNebula,Ricci2008THENEBULA}.

Proplyds are detected in strong \Ha\ emission, which serves as a signature of ionized gas near the ionization front, where UV radiation effectively ionizes the wind from the disk.
The \Ha\ surface brightness was first employed by \citet{Laques1979DetectionCamera} to estimate the electron density of the gas in proplyds, which at the time were proposed to be partially ionized globules. Subsequently, \citet{Henney1997OpticalOrion} introduced the first theoretical models, which reproduced the emission line profiles. Expanding on the latter study, \citet{Henney1999AProplyds} compared the models with high-resolution spectroscopy measurements of four proplyds acquired with the HIRES spectrograph on the Keck Telescope, and analyzed the spatial profiles of \Ha, H$\beta$, and forbidden lines, such as those of oxygen, sulfur, and nitrogen ions.
Forbidden lines have been the key tool to probe the density, temperature, and ionization state of the gas in proplyds \citep{Mesa-Delgado2012IonizedProblem}. Furthermore, forbidden lines of for instance, [Fe\,\textsc{ii}] and [S\,\textsc{ii}] have traced the supersonic jets and winds from proplyds \citep{Bally2001Irradiated1333,Kirwan2023AView}.

The characteristic cometary or teardrop shape of proplyds is created by the outflow of neutral gas from the disk as extreme- and far-UV energy photons heat the surface material and drive thermal winds. The neutral gas becomes photoionized as it expands and interacts with stellar UV radiation. The external photoevaporation of disk material leads to mass loss, estimated to be in the range from $10^{-8}$ to $10^{-6} M_\odot\,\text{yr}^{-1}$ for proplyds within 0.3 pc of $\theta^1$ Ori C, based on values inferred from observations \citep{Henney1998ModelingProplyds,Henney1999AProplyds, Henney2002Mass2} and theory \citep{Johnstone1998PhotoevaporationNebula,Storzer1999PhotodissociationOrion,Richling2000PhotoevaporationRadiation}. Recent 
ALMA Band 3 (3.1 mm) observations of a dozen proplyds in the ONC found values in the order of $10^{-7}$\,M$_\odot$\,yr$^{-1}$ \citep{Ballering2023IsolatingObservations}.

Submillimeter wavelength observations with ALMA confirm that the gas disks in the ONC are compact in comparison with the disks studied in environments where negligible UV radiation is present \citep{Mann2014ALMAPROPLYDS, Eisner2018ProtoplanetaryObservations, Boyden2020ProtoplanetaryALMA, vanTerwisga2023SurveySODA,Ballering2023IsolatingObservations}. Similarly, low disk dust masses are found in regions with weaker, but still relevant, UV radiation, such as the inner $\sim$0.5 pc of the $\sigma$-Orionis cluster \citep{Ansdell2017AnCluster,Mauco2023TestingMeasurements}. 
The extreme environment may dramatically alter the initial conditions for emerging planets, as the total disk mass limits the mass available for planet formation.

The majority of the past studies of proplyds were carried out with HST imaging in a limited number of filters, and spectroscopic analysis is available for only a few proplyds \citep[e.g.,][]{Henney1999AProplyds}. The Multi-Unit Spectroscopic Explorer (MUSE) is an integral field unit (IFU) spectrograph on the ESO Very Large Telescope (VLT). Used in the narrow-field mode (NFM), MUSE allows for the first time to both spatially and spectrally resolve the structure of proplyds in the ONC, as demonstrated by \citet{Kirwan2023AView} and \citet{Haworth2023TheBar}. The IFU data provides spectroscopy covering a wealth of emission lines at each position in the proplyds, and sub-arcsec spatial resolution of the objects.

Here, we present the morphology of 12 objects in seven emission lines, including \Ha, and forbidden oxygen, sulfur, and nitrogen lines. We measure the radius of the ionization front of proplyds, a key ingredient for estimating the mass-loss rate, in \Ha\ and forbidden oxygen lines. Using the forbidden oxygen lines in addition to the traditional tracer \Ha\ is important for gaining a more comprehensive understanding of the physical and chemical processes occurring in these systems. From the spectra, we also measure the stellar properties of the targets. 

The paper is structured as follows. In Sect. \ref{sect:observations}, we describe the MUSE observations and data reduction. The sample with the morphology of individual proplyds is presented in Sect. \ref{sect:present-sample}, and the outline of the analysis steps is given in Sect. \ref{sect:analysis}. We summarize the results in Sect. \ref{sect:results}, and further discuss them in Sect.~\ref{sect:discussion}.

\begin{table}
\begin{center}
\caption{Coordinates and projected separations of the detected proplyds.}
\begin{tabular}{l|ccc}
\hline \hline
Proplyd & RA & DEC & $d$(UV source) \\
 & hh:mm::ss.s & dd:mm:ss.s & [pc] \\
\hline
154-346 & 05:35:15.44 & $-$05:23:45.55 & 0.068 \\
167-325 & 05:35:16.72 & $-$05:23:25.5 & 0.009  \\
168-326 & 05:35:16.85 & $-$05:23:26.22 & 0.012  \\
170-249 & 05:35:16.96 & $-$05:22:48.51 & 0.068 \\
170-334 &  05:35:16.96 & $-$05:23:33.6 & 0.028 \\
170-337 & 05:35:16.97 & $-$05:23:37.15 & 0.031 \\
171-340 & 05:35:17.06 & $-$05:23:39.77 & 0.037 \\
173-236 & 05:35:17.34 & $-$05:22:35.81 & 0.095 \\
174-414 & 05:35:17.40 & $-$05:24:14.5 & 0.106 \\
177-341W & 05:35:17.66 & $-$05:23:41.00 & 0.049 \\
203-504 & 05:35:20.26 & $-$05:25:04.05 & 0.077 ($\theta^2$ Ori A)\\
244-440 & 05:35:24.38 & $-$05:24:39.74 & 0.06 ($\theta^2$ Ori A)  \\
& & &  0.31  ($\theta^1$ Ori C)\\
\hline
\end{tabular}
\tablefoot{The information is from \citet{OdellWen1994,Bally2000DisksNebula,Ricci2008THENEBULA,Mann2014ALMAPROPLYDS}. For the given projected separations $d$, the UV source is $\theta^1$ Ori C with the exception of 203-504 (irradiated by $\theta^2$ Ori A) and 244-440.}
\label{table:coordinates}
\end{center}
\end{table}
%\tablefoot
\begin{figure*}
    \centering
    \includegraphics[width=0.95\textwidth]{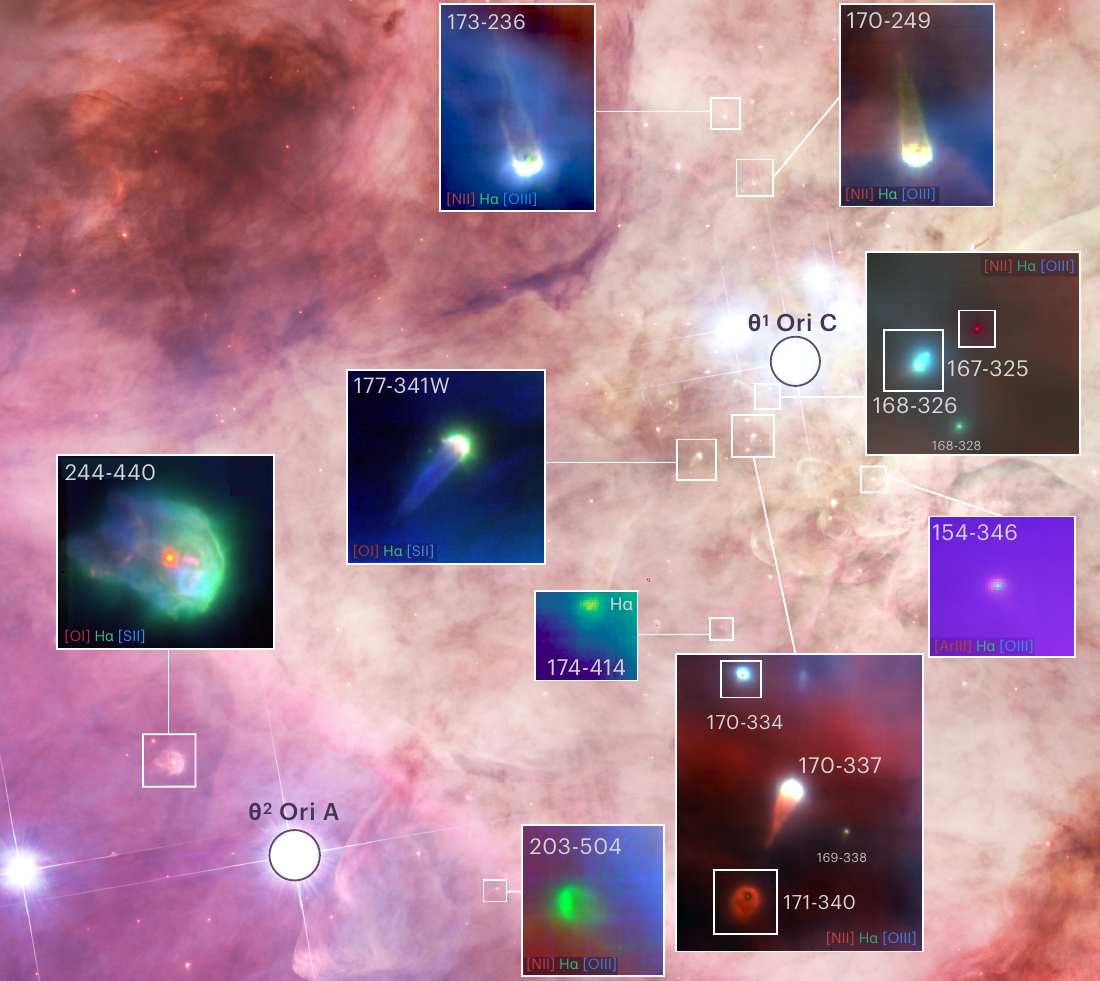}
    \caption{RGB images (insets) of continuum subtracted, single line integrated flux images of the sample acquired with MUSE. In each inset, a variety of emission lines are combined to highlight the morphology of the proplyd. The background image is a Hubble optical image of the Orion Nebula Cluster (Credit: NASA, ESA, M. Robberto (Space Telescope Science Institute/ESA) and the Hubble Space Telescope Orion Treasury Project Team). }
    \label{fig:thesample}
\end{figure*}

%--------------------------------------------------------------------
\section{Observations and data reduction}\label{sect:observations}
The targets in our sample were chosen to cover a large range of distances from $\theta^1$ Ori C, covering different levels of the incident radiation field and thus ionization. Furthermore, the targets were selected to have an overlap with ALMA observations, providing disk mass and size (\citealt{Mann2014ALMAPROPLYDS}, \citealt{Eisner2018ProtoplanetaryObservations}, \citealt{Boyden2020ProtoplanetaryALMA}). 

Observations were carried out with the MUSE integral-field spectrograph \citep{Bacon2010}, which is mounted on the Unit Telescope 4 (UT4) at the VLT. MUSE was operated in the Narrow Field Mode (NFM), which covers 4750--9350 \AA\ in a smaller field of view (FOV) of $\sim$7.5" $\times$ 7.5". On this FOV, the AO-correction is enhanced by the Laser Tomography Adaptive Optics (LTAO) mode leading to near diffraction limited images with a sampling of 0.025" per spatial pixel. In all the observations, the Natural Guide Star (NGS) was chosen to be the target, and was thus on-axis.

The observations presented here were carried out in three different observing programs between October 2019 and November 2022, always in Service Mode.
The observations for the proplyds 244-440 \citep[see][]{Kirwan2023AView}, 168-328, and 177-341W were taken in October 2019 (program ID 104.C-0963(A), PI: C. F. Manara), and those for the rest of the sample in January and February 2021 (program ID 106.218X.001, PI: C. F. Manara). 
Finally, we also include here the data taken in November 2022 during the Director Discretionary Time program 110.259E.001 (PI T. J. Haworth) for the proplyd 203-504, which are presented in \citet{Haworth2023TheBar}. 
MUSE was operated under clear sky conditions, and the full AO system was employed, using the stars at the center of our targets as NGS to achieve the best possible correction. The log file, including the airmass-corrected seeing and turbulence parameters are listed in Table \ref{Table:Obs-Log}.

The observations were generally composed by three dithered exposures. In each observation, the telescope was offset by 0.2" and 0.1" in the second and third exposures, respectively, and turned by 90 degrees with respect to the previous exposure to achieve the best possible corrections for bad pixels and instrumental effects. The exposure time in each exposure is reported in Table~\ref{Table:Obs-Log}.

Data reduction was carried out with the MUSE pipeline v2.8 \citep{Weilbacher2020TheInstrument} run in the ESO Reflex \citep{Freudling2013AutomatedAstronomy} environment. The pipeline carries out the classical data reduction steps, including bias, dark, flat computation and correction, wavelength calibration, flux calibration, sky subtraction, and exposure alignment and combination. The only steps that required non-standard settings were the alignment and combination of the cubes. Obtaining a satisfactory alignment was complicated due to only a small number of stars being present in the observed FOV. For that reason, some of the final cubes are obtained by combining a smaller set of dithering/rotation exposures (see Table~\ref{Table:Obs-Log}). Having a few stars available in the FOV can also decrease the accuracy of the coordinates, and the coordinates of the central stars are retrieved relative to those catalogued following the process described in Sect. \ref{subsect:I-front-measurement}.
% updated - \cfm{here we should mention that some final cubes are obtained combining a smaller set of dithering/rotation exposures, since the alignment was complicated.}
The final, wavelength calibrated data-cubes are in barycentric velocities, and the wavelength scale is in air wavelengths. The spectral resolution is R$\approx$2000--4000, corresponding to 170--90 km s$^{-1}$ from blue to red. We measure the spatial resolution of the images in the range FWHM$\approx$0.065--0.080" at $\sim$6760 \AA \, in our observations. %young disk-bearing stars 
\begin{figure*}
    \centering
    \includegraphics[width=0.99\textwidth]{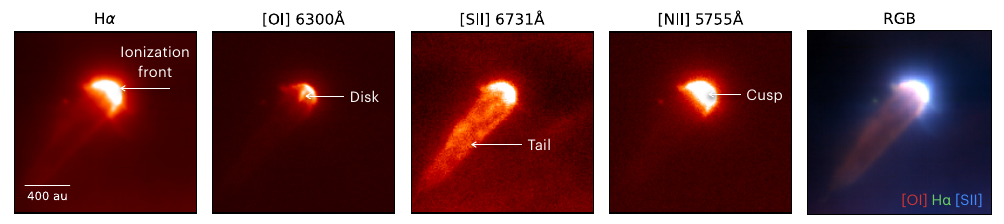}
    \caption{Continuum subtracted, single line integrated flux images of proplyd 177-341W. The panels show which parts of the system are visible in given emission lines.}
    \label{fig:morphology-example}
\end{figure*}

Continuum subtraction was then performed on sub-cubes spanning $\sim$100\AA\, to only include the spectral emission regions of interest. Each sub-cube was continuum-subtracted by fitting a second-order polynomial to the spectrum at each spatial pixel (spaxel) in the sub-cube.
This procedure was performed using the Python package MUSE Python Data Analysis Framework (\texttt{mpdaf}\footnote{https://github.com/musevlt/mpdaf}, \citealt{Piqueras2017MPDAFData}), in particular the function \texttt{spec.poly\_spec}. After this step, the sub-cubes were divided into smaller slabs with individual lines only, again using \texttt{mpdaf}.

\section{Morphology of observed proplyds}\label{sect:present-sample}
An overview of the ONC with the observed proplyds is presented in Fig. \ref{fig:thesample}. The view includes RGB images of nine proplyds: 154-346, 167-325, 170-249, 170-337, 173-236, 174-414, 177-341W, 203-504, and 244-440. In addition, there are proplyds in the background of several images: 168-326, 170-334, and 171-340. The coordinates of the targets are given in Table \ref{table:coordinates}. The lines associated with the RGB colors are chosen for each target to show the morphological characteristics described in this section.

A collage of each proplyd is shown in Fig.~\ref{fig-app:collage-1} and Fig.~\ref{fig-app:collage-2} in the following emission lines: \Ha, [O\,\textsc{i}] 6300\AA,  [O\,\textsc{ii}] 7330\AA,  [O\,\textsc{iii}] 5007\AA, [S\,\textsc{ii}] 6731\AA, [N\,\textsc{ii}] 5755\AA, and [N\,\textsc{ii}] 6584\AA. These lines are selected to show the various components of the systems, including the ionization front, the disk, the cusp, and the tail, as seen as an example based on proplyd 177-341W in Fig. \ref{fig:morphology-example}. The detailed analysis of these and other emission lines is deferred to forthcoming papers, while here we focus only on the measurement of the ionization front size, well traced by the H$\alpha$ and oxygen emission lines.

We first describe the morphology of the well-detected proplyds, targets for which the ionization front, the disk, and the tail can be distinguished. These targets appear with the classic teardrop shape and the elongated tail, covering $\sim$1.2-4" in the FOV, except for 244-440 which spans over $\sim$6.4". We then give an overview of the fainter objects, which cover $<$1" in the FOV, and the non-detections.

\subsection{Well-detected proplyds}
\subsubsection{170-249}
The proplyd 170-249 has an elongated tail in the north direction. This tail is visible in \Ha{}, forbidden sulfur species, and [N\,\textsc{ii}] 6584\AA{} lines. The cusp and the ionization front are bright in most of the emission lines, while the disk is traced in the [O\,\textsc{i}] 6300\AA\, and [S\,\textsc{ii}] 6312\AA\ lines. \\
In the FOV, the bright central region of a second object is visible in  [O\,\textsc{i}] 6300\AA\ and [N\,\textsc{ii}] 5755\AA\ lines to the southwest of 170-249 (not shown in the RGB image in Fig. \ref{fig:thesample}); no tail is distinguished. 
\subsubsection{170-337}
The proplyd 170-337 exhibits the elongated tail in the southeast direction, visible in [S\,\textsc{ii}] 6731\AA\ and [N\,\textsc{ii}] 6584\AA\ emission. The disk can be distinguished in \Ha\ and [O\,\textsc{i}] 6300\AA\ lines. The cusp and the ionization front are bright in most of the emission lines. \\
The FOV of 170-337 is rich with more targets:
\begin{itemize}
    \item In the southeast direction of 170-337, a well-resolved proplyd, designated 171-340, is visible with a round, extended cloud of ionized gas surrounding the disk. The proplyd differs from other objects by bearing a dark core, which may be a disk seen in silhouette against the background nebular light. The structure resembles the case of proplyd HST-10, also designated as 182-413 \citep{OdellWen1993}.\\
    A long tail is not visible, but the proplyd is extended in the direction opposite the UV source. The proplyd is bright in [S\,\textsc{ii}] 6731\AA\ and [N\,\textsc{ii}] 6584\AA\ lines. The central region appears bright in [S\,\textsc{ii}] 6731\AA.
    \item In the upper part of the cube in the northeast from 170-337, the cusp of the proplyd 170-334 is bright in most emission lines. The classic tail shape is visible in \Ha, and it is aligned with the one of 170-337. 
    \item There is a very small proplyd, 169-338, visible to the west of 170-337. We do not include it in the analysis as it is unresolved.
\end{itemize}
Although 171-340 and 170-334 are smaller than the central target 170-337, it is possible to measure their ionization front radius.
\subsubsection{173-236}
The proplyd 173-236 exhibits an elongated tail in the northeast direction. The star-disk region is traced by [O\,\textsc{i}] 6300\AA\ and [S\,\textsc{ii}] 6731\AA\ lines. With a small, bright central part visible in \Ha. 
There is a bubble-like structure southeast from the center of the FOV, which is most prominent in [S\,\textsc{ii}] 6731\AA\ emission (not shown in Fig. \ref{fig:thesample}).
\subsubsection{177-341W}
The RGB image of proplyd 177-341W, shows both a bright cusp and a highly elongated tail in the southeast direction, which is brightest in [S\,\textsc{ii}] and [N\,\textsc{ii}] emission lines. Each emission line traces the cusp with the ionization front. The disk can be distinguished in the [O\,\textsc{i}] 6300\AA\ emission line, oriented perpendicular to the cusp, in a northwest to southeast direction.

\subsubsection{244-440}
Proplyd 224-440 is one of the largest in the ONC, spanning about 6.4" from the tip of the longest tail to the ionization front. The tail structure is more complex compared to the other proplyds, resembling a tooth-like shape with two to three tail components depending on the observed wavelength; a JWST/NIRCam view is shown by \citet{Habart2023PDRs4AllNebula}.\\
The morphology is striking with different emission lines tracing a variety of the system's components: the cusp and the tail being visible in different ionization states of forbidden oxygen, sulfur, and nitrogen; the disk is bright in [O\,\textsc{i}] 6300\AA.
The jet is visible in the MUSE data in the [O\,\textsc{i}] 6300\AA\ and various [Fe\,\textsc{ii}] lines \citep{Kirwan2023AView}. 

\subsubsection{203-504}
The proplyd 203-504 is coincident with the Orion Bar, which lies approximately 112'' to the southeast of the Trapezium. The teardrop shape is less elongated and more oval compared to the targets close to $\theta^1$ Ori C, and the morphology is bright in \Ha{} and [N\,\textsc{ii}] 5755\AA{}. The cusp is brightest in \Ha{}, [N\,\textsc{ii}] 5755\AA{}, while the disk region is traced in the forbidden oxygen lines.
\citet{Haworth2023TheBar} discusses the target in depth. This is the only proplyd in our sample which is directly irradiated only by $\theta^2$ Ori A.

\subsection{Faint proplyds}
\subsubsection{154-346}
The star-disk region of proplyd 154-346 is visibly bright in each emission listed above, except for [O\,\textsc{iii}] 5007\AA\ and [N\,\textsc{ii}] 5755\AA. The proplyd is very small and the classic drop-like shape can be distinguished faintly. 
\subsubsection{167-325 and its FOV}
The proplyd 167-325 is closest to $\theta^1$ Ori C  (0.009 pc) in our field. Only the star-disk region of the system is visible, traced in [N\,\textsc{ii}] 6584\AA\ line and faintly in forbidden sulfur species. 
In the same FOV, there are three objects:
\begin{itemize}
    \item Two proplyds appear very close to each other in the southeast direction of the central target; the classic shape of the foreground proplyd (168-326) is visible while only a part of the background proplyd can be distinguished. The tail of 168-326 is visible in [N\textsc{iii}] 6584\AA\ line. The star-disk region is traced by \Ha, [O\,\textsc{ii}] 7320\AA\,, 
 [O\,\textsc{iii}] 5007\AA\,, and [S\,\textsc{ii}] 9068\AA\, lines. 
\item There is another proplyd in the south of 167-325 and 168-326. It is much smaller, and its ionization front size is not measured.
%but the tail is visible in [N\,\textsc{ii}] 6584\AA\ line and the star-disk region is visible in the same lines as 169-327.
\end{itemize}

\subsubsection{174-414}
The proplyd 174-414 appears at the very edge of the cube which was planned to be an observation of HST-10. The teardrop shape is visible faintly in \Ha{} and [N\,\textsc{ii}] 6584\AA{}. The central part is bright in [O\,\textsc{i}] 6300\AA{}, [O\,\textsc{ii}] 7320\AA\, and forbidden nitrogen and sulfur lines.  

\subsection{Non-detections}
The proplyds 142-301, 167-231, and HST-10 were targeted on our MUSE programs, but not detected. \\
In the case of proplyd 159-350, the pointing was not accurate and the observation of this proplyd was not acquired; instead, the pointing was centered at the nearby proplyd 154-346.

\begin{figure}[]
%    \centering
\includegraphics[width=.99\columnwidth]{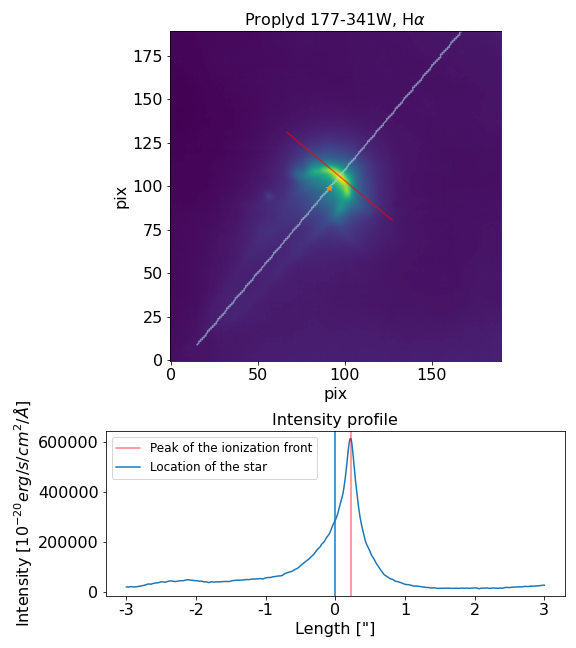}
    \caption{Method for measuring the ionization front radius of a proplyd. Top panel: the intensity cut is taken along the diagonal line. The red line marks our identification of the ionization front. Bottom panel: intensity cut along the diagonal line marked in the top panel.}
    \label{fig:measurement-method}
\end{figure}

\begin{table*}
\begin{center}
\begin{threeparttable}
\caption{Measured values for the observed proplyds}
\begin{tabular}{l|cccc|c|l}
\hline \hline
Proplyd & R$_{IF}$ \Ha{} & R$_{IF}$[O\,\textsc{i}]  & R$_{IF}$ [O\,\textsc{ii}] & R$_{IF}$ [O\,\textsc{iii}] & $\dot{M}_{\rm loss}$ \Ha{} & M$_{\rm disk}$ \\
 &  ["] & ["] & ["] & ["] & M$_\odot\,yr^{-1}$ & [$M_{\rm Jup}$] \\
\hline
154-346 & 0.10* &  0.11* &   0.10* &  \nodata &   1.3 $\times$ 10$^{-7}$ & \nodata\\

167-325 & 0.10* &  0.10* &   \nodata &     \nodata &   8.57 $\times$ 10$^{-7}$ & \nodata \\

168-326 & 0.10 $\pm$ 0.01 &  0.10 $\pm$ 0.01 &   0.10 $\pm$ 0.01 &     0.12 $\pm$ 0.01 &   6.43 $\times$ 10$^{-7}$ &   5.0 $\pm$ 3.0\\

170-249 & 0.23 $\pm$ 0.02 & 0.17 $\pm$ 0.02 &    0.21 $\pm$ 0.02 &     0.29 $\pm$ 0.03 &   4.6 $\times$ 10$^{-7}$ &  27 $\pm$ 0.4\\

170-334 &  0.12* &   0.09* & 0.07* &     0.10*&   3.9 $\times$ 10$^{-7}$ &  \nodata  \\

170-337 &  0.17 $\pm$ 0.02 &   0.12 $\pm$ 0.01 & 0.14 $\pm$ 0.01 &     0.17 $\pm$ 0.02 &   5.5 $\times$ 10$^{-7}$ & 7.13 $\pm$ 1.08 \\

171-340 &  0.19 $\pm$ 0.02 & 0.16 $\pm$ 0.02 &   0.18 $\pm$ 0.02 &     0.18 $\pm$ 0.02 &   5.46 $\times$ 10$^{-7}$ &  15.83 $\pm$ 0.42 \\

173-236 & 0.23 $\pm$ 0.02 &   0.20 $\pm$ 0.02 &  0.23 $\pm$ 0.02 &     0.24 $\pm$ 0.02 &   2.85 $\times$ 10$^{-7}$ & 45 $\pm$ 1.0\\

174-414 & 0.13* &   0.10* &  0.15* &    \nodata &   1.07 $\times$ 10$^{-7}$ &    \nodata \\

177-341W & 0.2 $\pm$ 0.02 &  0.18 $\pm$ 0.02 &   0.21 $\pm$ 0.02 &     0.26 $\pm$ 0.03 &   5.11 $\times$ 10$^{-7}$ &  7.48 $\pm$ 0.45 \\

203-504 & 0.21 $\pm$ 0.02  & 0.18 $\pm$ 0.02 &   0.23 $\pm$ 0.05 &     \nodata &   2.5 $\times$ 10$^{-7}$ &    \nodata \\

244-440 &  2.0 $\pm$ 0.1 &   1.95 $\pm$ 0.0975 & 1.97 $\pm$ 0.098 &     2.18 $\pm$ 0.22 &   9.45 $\times$ 10$^{-6}$ &\\
\hline
\end{tabular}
\tablefoot{* Uncertain value, considered as lower limit.
The values of disk dust masses have been estimated by \citet{Mann2014ALMAPROPLYDS} and \citet{Eisner2018ProtoplanetaryObservations}, except for 170-249 and 173-236 \citep{Ballering2023IsolatingObservations} (T=20 K).}
\label{table:proplyd-parameters}
\end{threeparttable}
\end{center}
\end{table*}
%\tablefoot
% [O\,\textsc{i}] 6300\AA\,, [O\,\textsc{ii}] 7330\AA\,, and [O\,\textsc{iii}] 5007\AA.

\section{Analysis}\label{sect:analysis}

\subsection{Measurement of the ionization front radius}\label{subsect:I-front-measurement}
Our goal is to study the radius of the ionization front (I-front) as a function of the incident radiation field. Here, we focus on measuring the I-front radius for four lines: \Ha, [O\,\textsc{i}] 6300\AA\,, [O\,\textsc{ii}] 7330\AA\,, and [O\,\textsc{iii}] 5007\AA. The \Ha\ line allows us to detect the region where ionized gas transitions to neutral gas; this has been used as the tracer of the ionization front in past studies \citep[e.g.,][]{Henney1998ModelingProplyds, Henney1999AProplyds}. The forbidden oxygen lines allow us to additionally trace the ionization structure for different ionization states.

We located the I-front at the bright cusp on the side closer to the UV source. The main UV influx comes from $\theta^1$ Ori C for all proplyds except 203-504 and 244-440, which appear to be primarily irradiated by $\theta^2$ Ori A \citep{ODell2017WhichNebula,Haworth2023TheBar}.

As the coordinates of the stars in the MUSE cubes might not directly correspond to those cataloged because of poor astrometric calibrations of our dataset (described in Sect. \ref{sect:observations}), we determined the coordinates of the central stars in the MUSE data directly. 
We created a sub-cube with continuum emission only, that is, a region free of lines, at 6740-6780\AA. We 
%continue with a smaller sub-image around the coordinates of the star reported in the literature to exclude other sources, and 
located the peak of the continuum emission by using the \texttt{mpdaf} function \texttt{peak} in the region of the star. The coordinates of the peak were used as the location of the star for the following steps and the following measurements were always carried out using relative distances (Fig. \ref{fig:measurement-method}).
 
We then retrieved a radial cut of the proplyd intensity profile in each of the investigated lines by tracing a  6'' line centered at the coordinates of the central star. We set the position angle of the radial cut in the direction between the central star and the peak of the emission in the ionization front, which in turn is in the direction of the ionizing star in all cases, with the exception of 244-440 and 203-504. For 203-504, the angle between the proplyd and $\theta^2$ Ori A was calculated based on their catalogued coordinates.
The intensity profile along the line was extracted by using cubic interpolation with the \texttt{scipy} function \texttt{map$\_$coordinates}.
The location of the I-front was determined relative to the line's center, the $\sim$0'' mark on the x-axis of the intensity profile. For the majority of the sample, an intensity peak corresponding to the I-front, a bright region in the direction of the UV source, lies at a small offset from the center. This offset is the radius of the I-front.

The I-front radius value measured from the intensity profile overlaps roughly with the brightest area visible on the image of each proplyd; the scale parameters of the image can slightly change the boundary of the brightest area. An example of the measurement method is shown in Fig. \ref{fig:measurement-method}.

We estimated the uncertainty by repeating the measurement for two more position angles ($\pm$15$^{\circ}$), and calculating the percentage change between the values of I-front radii. The typical uncertainty in the cases of good detection is 10\% (see Table \ref{table:proplyd-parameters}).
 
Proplyds 154-346, 167-325, 170-334, and 174-414 make up a much smaller part in the FOV compared to the well-detected targets. In those cases, no distinct peak is visible in the intensity profile and we estimated the I-front's position based on locating the outer part of the bright center of the target. This is the lower limit of the I-front.

The identification of the I-front was straightforward for the \Ha\ lines of each well-detected proplyd. The I-front was fainter and less clear to distinguish the oxygen lines of several proplyds. There were two intensity peaks for the oxygen lines compared to the single peak in \Ha\ intensity profiles: the higher peak coincided with the star's location, and the smaller peak coincided with the I-front. The principle of the measurement remained the same. The measured I-front radii in the various lines are reported in Table \ref{table:proplyd-parameters}.

\begin{figure*}[h]
\centering
    \includegraphics[width=0.8\textwidth]{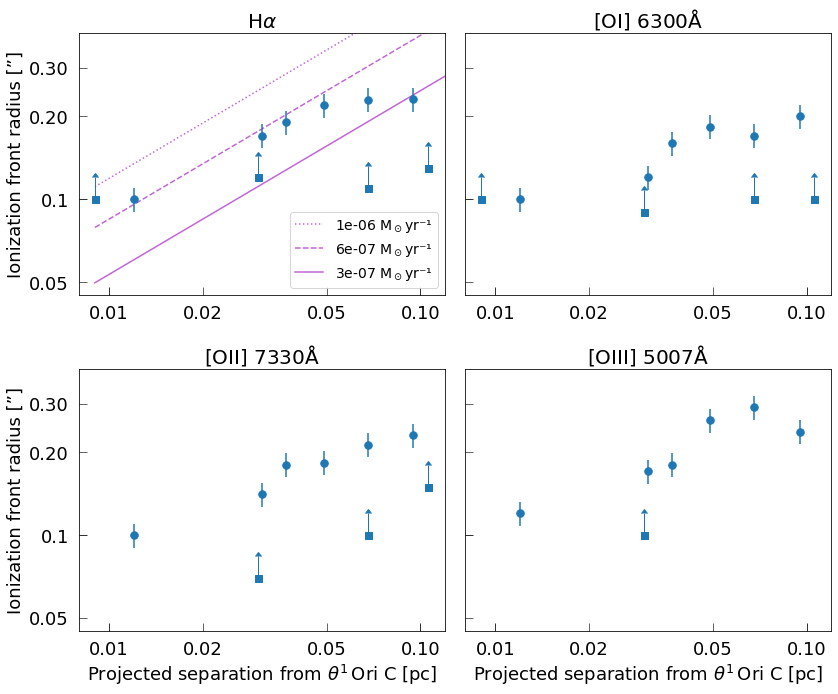}
      \caption{Ionization front radius versus the projected separation from $\theta^1$Ori C. Squares mark proplyds for which the I-front measurement has a higher uncertainty. The radii of the ionization fronts increase with larger projected distances from the UV source. Due to noisy subcubes, measurements were not taken for proplyd 167-325 (closest to the UV source) in the bottom panels. For the same reason, the panel of [O\,\textsc{iii}]5007\AA{} excludes the radii of several targets.}
    \label{fig:I-frontSizes}
\end{figure*}

\begin{table}
\begin{center}
\caption{\label{table:star_pars} Stellar parameters derived for the sampled proplyds}
\begin{tabular}{l|cc|ccc}
\hline \hline
Proplyd & SpT & $A_V$ & $T_{\rm eff}$ & 
$\log(L_\star)$ & $M_\star$ \\
& & [mag] & [K] & [$L_\odot$] & [$M_\odot$] \\
\hline
% 154-346 & - & & & & & no spec? \\ Instead of 159-350, we have 154-346.
154-346 & M0 & 4.5 & 3900 & 0.46 & 0.60  \\
167-325 & \nodata &\nodata &\nodata &\nodata & \nodata \\ 
168-326 & \nodata &\nodata &\nodata &\nodata & \nodata \\ 
% NB! Mistakenly saved as 168-326 in files
 % actually  & & & &\\
170-249 & M4.5 & 3.5 & 3085 & -0.48  & 0.20 \\
170-334 &  K5.5 & 3.0 & 4162 & 1.05 & 1.02  \\
170-337 & K6 & 3.5 & 4115 & 0.39 & 0.79  \\
171-340 &  M1 & 3.0 & 3720 & 0.22 & 0.47  \\
173-236 & K5.5 & 5.0 & 4160 & 0.96 & 0.99 \\
174-414 & M5 & 6.0 & 2980 & -0.34 & 0.18 \\
177-341W &  K5.5 & 4.3 & 4160 & -0.30 & 0.91\\
203-504 & K5.5 & 3.0 & 4160 & -0.74 & 0.72 \\
244-440 & M0 & 3.0 & 3900 & 0.42 & 0.59  \\
\hline
\end{tabular}
\end{center}
\end{table}

\subsection{Stellar parameters}\label{sect::star_properties}

The MUSE data allow us to determine the stellar parameters of the central objects for the observed proplyds. 
We extracted the spectra of the stars using a circular aperture of 0.05"--0.075" depending on the target and an annulus of 0.075"--0.175" to estimate the sky emission. The spectra for 167-325 and for 168-326 could not be extracted as the S/N ratio of the continuum is too low.

The spectra were then compared with the set of pre-main-sequence empirical templates from \citet{Manara2013X-shooterObjects, Manara2017AnStars}, and Claes et al. (in prep.) to determine the spectral type (SpT) and extinction ($A_V$). The latter was determined assuming the reddening law by \citet{Cardelli1989TheExtinction} and a value $R_V$=3.1. Typical uncertainties on these derived parameters are of 1-2 sub-classes for the SpT and of 0.5 mag on $A_V$, but they depend on the quality of the spectrum in the redder part ($\lambda>$750 nm) which is sometimes suboptimal for several reasons, including strong emission lines and telluric absorption. The SpT was converted into effective temperatures ($T_{\rm eff}$) using the relation by \citet{Herczeg2014ANPROPERTIES}. From the $J-$band magnitude \citep{Robberto2010ANEAR-INFRARED}, we then estimated the stellar luminosity ($L_\star$) using the bolometric correction by \citet{Herczeg2014ANPROPERTIES}, to finally derive a stellar mass ($M_\star$) using the evolutionary tracks by \citet{Siess2000AnStars}. The latter was used because most of the targets appear to be located on the Hertzsprung-Russel Diagram above the 1 Myr isochrone, not covered by other tracks, as expected due to their young age. 

The derived parameters are reported in Table~\ref{table:star_pars}, and the best fits are shown in Fig.~\ref{fig:best_fit_spt-1}. We compare the derived values with \citet{Fang2021AnCluster}, but only a sub-sample of our targets are covered by that study, and they show a good agreement within one subclass with our estimates.

\section{Results}\label{sect:results}

\subsection{Ionization front radius}
The ionization front radius of proplyds enables us to infer the mass-loss rate of proplyds, which is a critical factor in the evolution of protoplanetary disks. %We have measured the ionization front (I-front) radii ($R_{IF}$) of 12 proplyds in the ONC.
The radii of targets irradiated by $\theta^1$ Ori C are shown as a function of the projected distance from the massive star in Fig. \ref{fig:I-frontSizes}.  We note that given the distance to the ONC and the generally high optical extinctions towards the region, the \textit{Gaia} parallaxes for our targets are not always available, or they are too uncertain to provide reliable individual distances to the targets. For this reason, we assume a common distance of 400 pc for the targets in our sample, and only consider projected separations from the ionizing stars in our analysis.
% with values spanning 0.1-0.23'' (40-92 au) for \Ha\ and 0.1-0.28'' (40-115.9 au) for forbidden oxygen lines over 0.012--0.106 pc.
In the panel of \Ha{} emission, we compare the measurements with predicted values for a set of mass-loss rates, as described in Sect. \ref{subsect:massloss}.

\begin{figure}[]
    \centering
    \includegraphics[width=0.99\columnwidth]{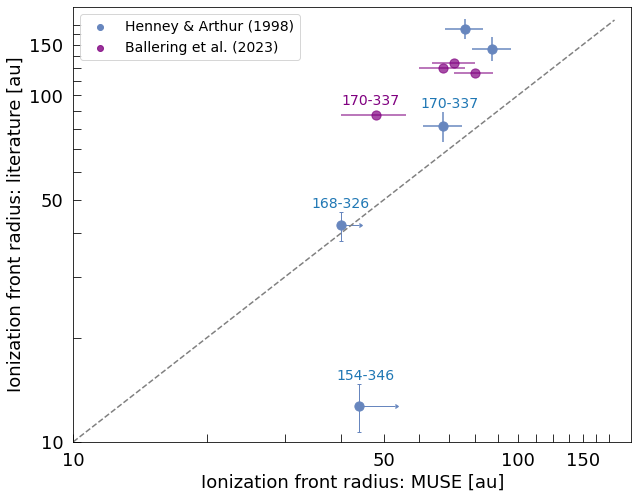}
    \caption{Ionization front sizes from literature \citep{Henney1998ModelingProplyds,Ballering2023IsolatingObservations} versus as measured with MUSE IFU data. We note that the measurement methods differ slightly from each other.}
    \label{fig:lit_Ifront}
\end{figure}

\begin{figure*}[]
\centering
    \includegraphics[width=0.8\textwidth]{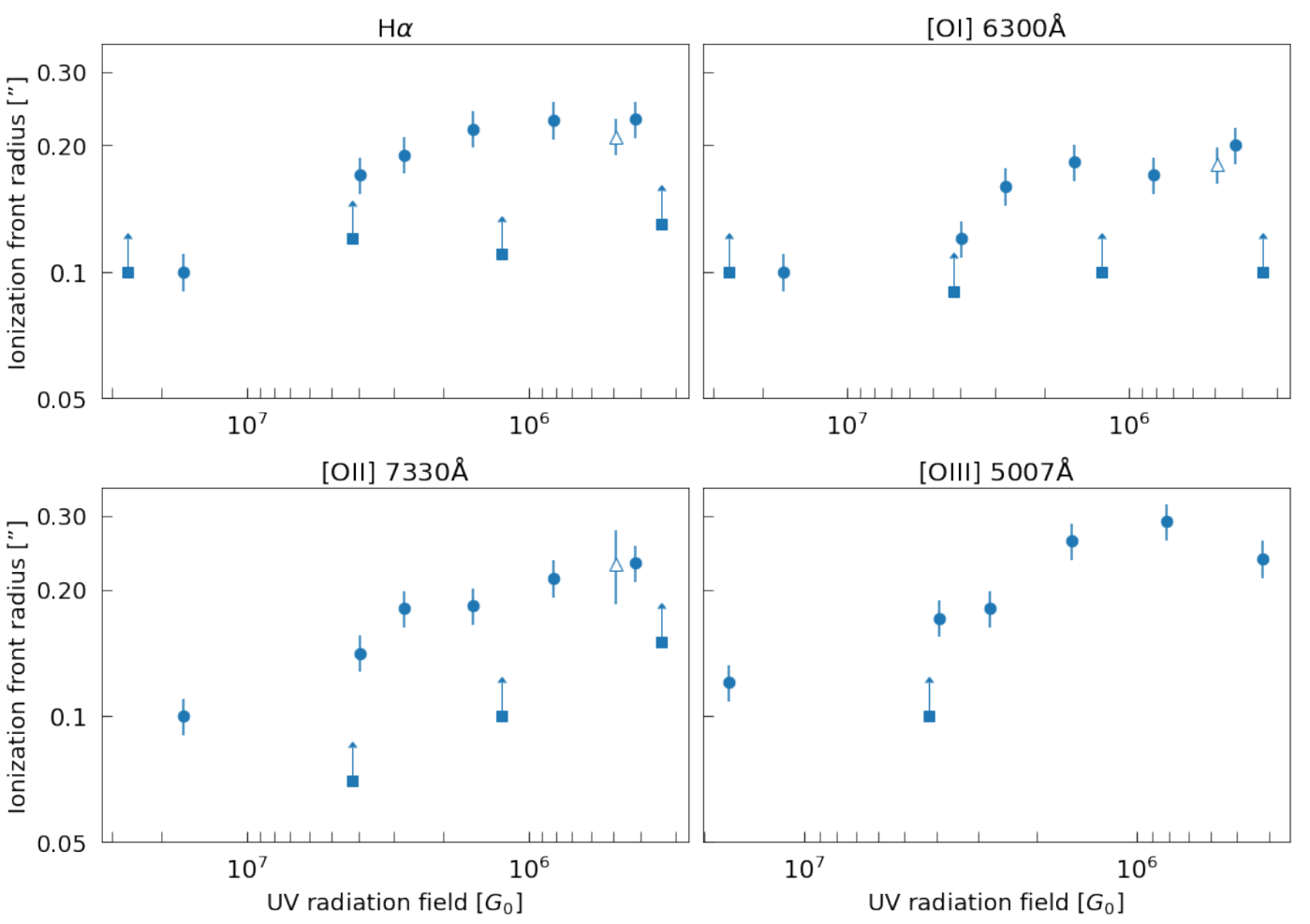}
     \caption{Ionizing radiation strength as a function of the ionization front radius. The diamond marks proplyd 203-504, irradiated mainly by  $\theta^2$ Ori A. For the rest of the targets (dots and squares), the UV source is $\theta^1$ Ori C. Squares mark proplyds for which the I-front measurement has a higher uncertainty. 244-440 is not included in this figure due to the stark difference in size compared to the rest of the sample.}
    \label{fig:G0-vs-ifront}
\end{figure*}

Figure \ref{fig:I-frontSizes} shows that the I-front radius is generally larger at lower ionization rates, that is, at larger projected distance, in all tracers discussed here. The proplyds with more uncertain measurements do not however typically follow the trend. The proplyds with more uncertain measurements are small and faint compared to the rest of the sample, and the I-front measurement could not be carried out with the same approach.\\
To explore the significance of the relation between I-front radii ($R_{IF}$) and projected distance ($d_p$), we carried out two analyses similarly to \citet{McLeod2021TheEnvironment}. Firstly, we tested the null-hypothesis that there is no relation between the parameters. We used a bootstrap sampling method with 10$^4$ iterations, where we randomized the x-axis values and computed the Spearman correlation coefficient ($r_s$). By calculating the $\sigma$ distribution of the data, we can exclude the null-hypothesis with a larger than 2$\sigma$ confidence in all cases (p<0.05).
Secondly, we assessed the uncertainty of the correlation coefficient by bootstrapping with 10$^4$ iterations while varying the y-axis values within their respective uncertainties. We adopted a truncated Gaussian distribution to model the uncertainty for the lower limits values in the $R_{IF}$ measurements. In those cases, the bootstrapping was selected within this truncated Gaussian distribution with a standard deviation of 20\% of the lower limit value. Both of these tests were then compared with the actual coefficient $r_s$. We also report the p-value of the null hypothesis randomizing the x-axis values. The p-value was calculated as the number of correlation coefficients larger (for positive correlation) than 1$\sigma$ of bootstrapped observations, and divided by the number of iterations. We excluded proplyd 244-440 from the given test and those which follow.

We found p-value~$\simeq$~0.02 and $r_s\simeq$ 0.62, indicating a positive relation (Fig. \ref{fig-app:correlation}a).

The sub-cube was too noisy for the measurement of $R_{IF}$ in the case of 167-325 at [O\,\textsc{ii}] 7330\AA{} and [O\,\textsc{iii}] 5007\AA{}, and for proplyds 174-414 and 203-504 at [O\,\textsc{iii}] 5007\AA{}; they are not included in Fig. \ref{fig:I-frontSizes}.

In Fig.~\ref{fig:lit_Ifront}, we show the comparison between our measurements of $R_{IF}$ and those from the literature. The I-front radii of proplyds 154-346, 168-326, 170-337, 170-334, and 177-341W were derived by \citet{Henney1998ModelingProplyds}, who fit models to observed \Ha\,intensity profiles.
Proplyd 168-326 is within the uncertainties of the literature values; proplyd 154-346 is a factor $\sim$3.5 lower in the literature, and the rest are higher. The differences in the measured values could be explained by the measurement approaches.\\
\citet{Vicente2005SizeCluster} measured the "chord diameters" of a sample of proplyds, in which 244-440 is common with our sample. The chord diameter is defined as the diameter of a circle fitted to the contour of the proplyd cusp boundary, plus $\sim$30\% the average background intensity. This approach places the I-front radius higher than our value, roughly at 2.8" or 1120 au; the measurement is not shown in Fig.~\ref{fig:lit_Ifront} due to the great difference in size compared to the rest of the targets.

Recently, \citet{Ballering2023IsolatingObservations} estimated the I-front radii of proplyds 170-249, 170-337, 173-236, and 177-341W marking the disk center and the outer edge of the ionization front on surface brightness cuts of ALMA Band 3 images. Our values are systematically lower compared to the ALMA study, because the I-front marker was positioned past the peak's inflection point, not on the peak.

\subsection{Ionization front radius and incident radiation field strength}

\begin{figure*}[h]
\centering
    \includegraphics[width=0.8\textwidth]{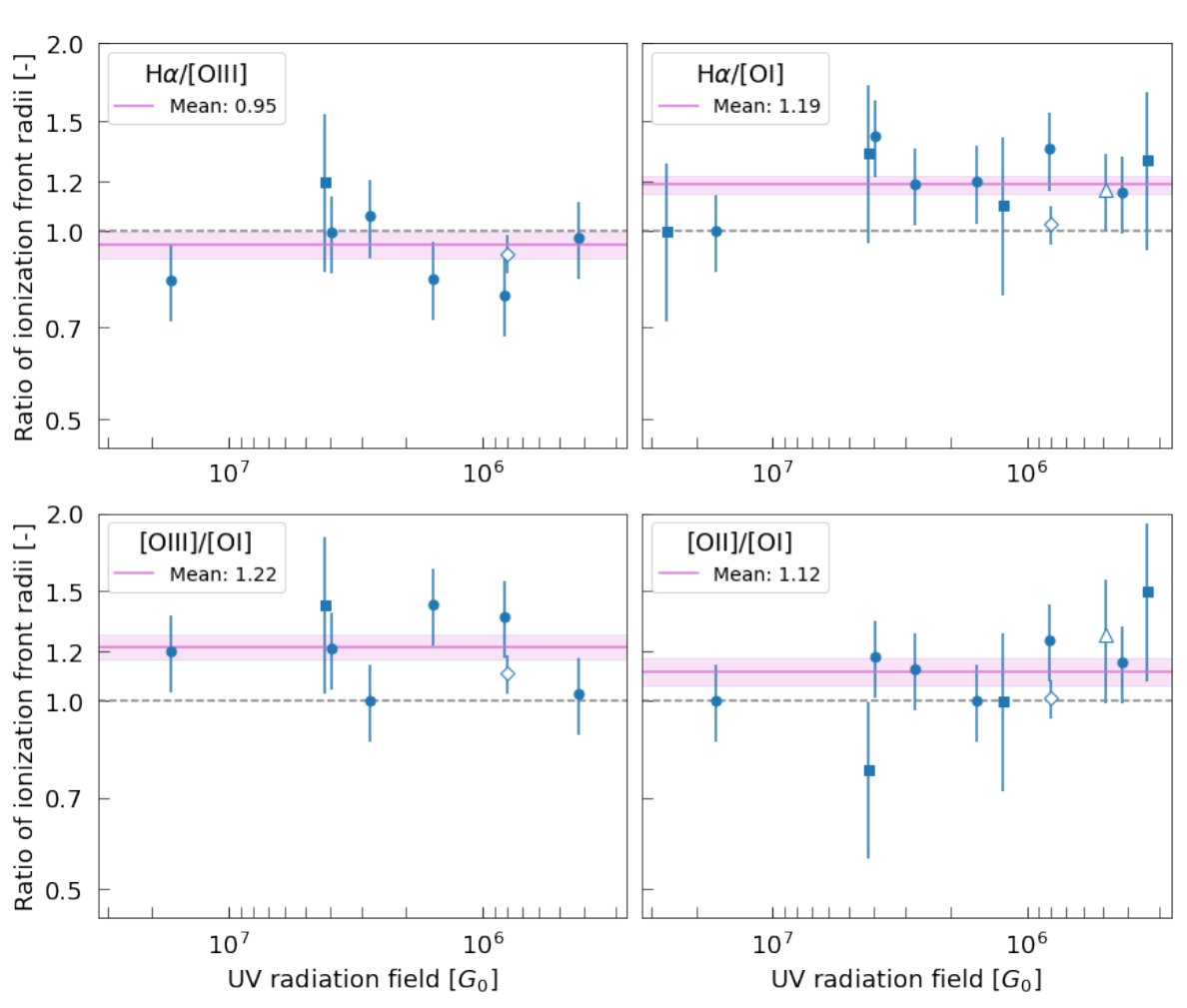}
    \caption{Ratios of ionization front sizes of H$\alpha$, [O\,\textsc{iii}], [O\,\textsc{ii}], and [O\,\textsc{i}] as a function of the incident UV field. The empty diamond marks 244-440, and the empty triangle marks 203-504. The mean and SEM are marked in violet.}
    \label{fig:O-ratios}
\end{figure*}

The strength of the FUV environment is described in the units of the Habing radiation field; 1 $G_0 = \rm 1.6 \times 10^{-3}\, erg\, cm^{-2}\, s^{-1}$ \citep{Habing1968TheA}. We calculate the strength of the incident radiation field using the following adimensional parameter:
\begin{flalign}
& F_{FUV} = \frac{1}{G_0}\frac{L_\mathrm{FUV}}{4\pi r^2},
\label{eq:G0}
\end{flalign}
 where the far-UV (FUV) luminosity $L_\mathrm{FUV}$ is derived from the values of FUV radiation reported as a function of stellar mass by \citet{WinterHaworth2022}. The stellar mass was considered to be 45 M$_{\odot}$ for $\theta^1$ Ori C and 39 M$_{\odot}$ for $\theta^2$ Ori A \citep{Simon-Diaz2006DetailedNebula}. The ionizing radiation strength for our sample spans from $10^5$--$10^7$ $G_0$ over projected distances of 0.009--0.31 pc. These values are consistent with the literature: the FUV field incident upon most of the proplyds in the vicinity of $\theta^1$ Ori C is estimated to be $F_{FUV}$ $\geq 10^6$. For 244-440, the value is estimated to be on the order of $10^{5} G_0$ assuming irradiation by $\theta^2$ Ori A, which agrees with the value found by \citet{Vicente2009PhDThesis}. However, \citet{ODell2017WhichNebula} note that the proplyd may be ionized by both \thetaonec{} and $\theta^2$ Ori A, which would increase the radiation strength.

In Fig. \ref{fig:G0-vs-ifront}, the I-front radii of the proplyds irradiated by $\theta^1$ Ori C, as well as proplyd 203-504 irradiated by $\theta^2$ Ori A, are plotted as a function of the incident UV radiation field. The I-front radii are larger at lower incident fields. We repeat the statistical test described above, and find $r_s\simeq$ -0.65, indicating a negative relation (Fig. \ref{fig-app:correlation}b), and p-value~$\simeq$~0.02.\\
The proplyd 203-504, irradiated by $\theta^2$ Ori A, follows very well the trend of the proplyds irradiated by $\theta^1$ Ori C. The proplyd 244-440 appears to be much larger than any other proplyd, and it is not included in the figure. 

To compare the different levels of ionization, we show the ratios of $R_{IF}$ for the H$\alpha$ line and for different forbidden oxygen lines as a function of the UV field in Fig. \ref{fig:O-ratios}. We calculate the mean and its standard deviation for the following ratios.
The ionization front radius increases for tracers that are more ionized: 
%R$_{IF}$[O\,\textsc{iii}]/ R$_{IF}$[O\,\textsc{i}] > R$_{IF}$[O\,\textsc{ii}]/ R$_{IF}$[O\,\textsc{i}]. Similarly,
R$_{IF}$[O\,\textsc{iii}]/ R$_{IF}$[O\,\textsc{i}] $\simeq$ 1.22$\pm$0.06 $>$ R$_{IF}$[O\,\textsc{ii}]/ R$_{IF}$[O\,\textsc{i}] $\simeq$ 1.12$\pm$0.06. We see an increasing trend as expected, but the statistical significance is limited by the small sample size. More data would be needed for a firm confirmation. The same applies to the following case: R$_{IF}$H$\alpha$/R$_{IF}$[O\,\textsc{iii}] $\simeq$ 0.95$\pm$0.05.

\subsection{Mass-loss rate}\label{subsect:massloss}

In Fig. \ref{fig:I-frontSizes} (panel of \Ha{}), the measured values of I-front radii are compared with predicted ionization front radii based on a set of mass-loss rates described by the following equation from \citet{WinterHaworth2022}: 

\begin{multline}
    r_\mathrm{IF} \approx 1200 \, \left( \frac{\dot{M}_\mathrm{loss}}{10^{-8} \, M_\odot \, \mathrm{yr}^{-1}}\right)^{2/3} \\ 
    \times \left( \frac{\dot{N}_\mathrm{Ly}}{10^{45} \, \mathrm{s}^{-1}}\right)^{-1/3} \left( \frac{d}{1\,\mathrm{pc}} \right)^{2/3} \, \mathrm{au},
    \label{eq:RIFMdot}
\end{multline} 

where $\dot{M}_{\rm loss}$ is the mass-loss rate of the proplyd, $d$ the projected separation from an irradiation source emitting $\dot{N}_{\rm Ly}$ ionizing photons per second.
For $\theta^1$ Ori C, we use $\dot{N}_{\rm Ly}=1.60\times 10^{49}$ photons s$^{-1}$, and $\dot{N}_{\rm Ly}=1.08\times 10^{49}$ photons s$^{-1}$ for $\theta^2$ Ori A, based on the values of stellar EUV luminosity as a function of stellar mass from \citet{WinterHaworth2022}. 

In the panel of \Ha{}, of Fig.~\ref{fig:I-frontSizes}, curves for three different mass-loss rate values are partly overlapping with the data points, with $\dot{M}_{\rm loss}$ decreasing over larger projected distances, as expected from theoretical models of disk photoevaporation \citep[e.g.,][]{Johnstone1998PhotoevaporationNebula, Storzer1999PhotodissociationOrion}. Four of the proplyds are at the curve marking $6\times10^{-7}$\,M$_\odot$\,yr$^{-1}$, while the rest of the proplyds vary slightly more, falling between $1\times10^{-6}$ and $3\times10^{-7}$$ $\,M$_\odot$\,yr$^{-1}$. The inferred $\dot{M}_{\rm loss}$ values are reported in Table \ref{table:proplyd-parameters}.

\begin{figure}[]
    \centering
\includegraphics[width=\columnwidth]{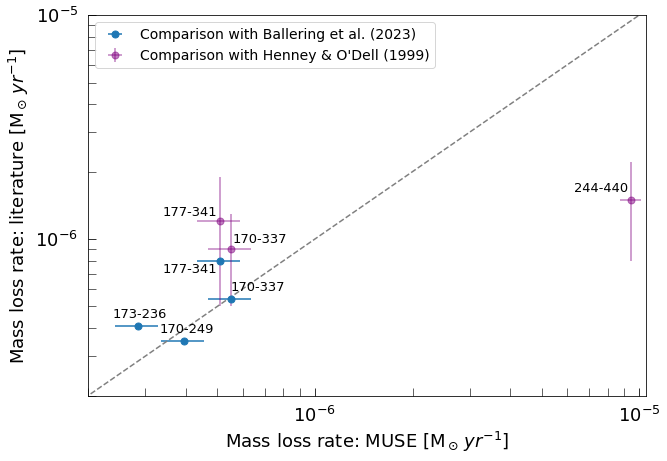}
    \caption{Mass-loss rates of MUSE compared with the values reported by \citet{Henney1999AProplyds,Ballering2023IsolatingObservations}.}
    \label{fig:Mloss-comparison}
\end{figure}

The comparison of our estimates of $\dot{M}_{\rm loss}$ with values reported in the literature is shown in Fig.~\ref{fig:Mloss-comparison}. The values are very consistent, except for 244-440 which has the highest $\dot{M}_{\rm loss}$ value in our sample, $\sim8.06\times10^{-6}$\,M$_\odot$\,yr$^{-1}$.
We discuss the values of $\dot{M}_{\rm loss}$ and the comparison with literature further in Sect. \ref{subsect-discussion:massloss}, and the case of 244-440 in Sect. \ref{sect:discussion-244-440}.

\begin{figure}[]
    \centering    \includegraphics[width=\columnwidth]{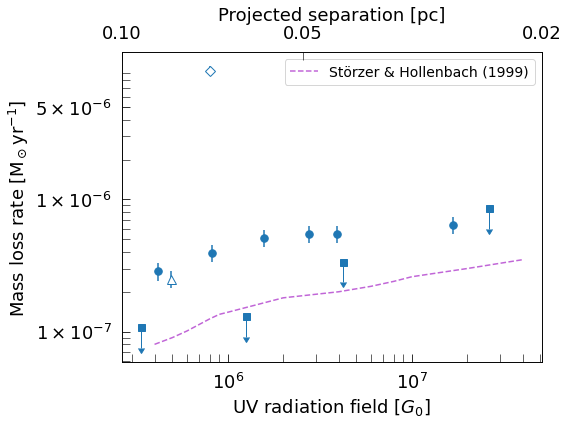}
    \caption{Mass-loss rate vs incident radiation field. The dots represent well-detected proplyds, and the squares denote the proplyds with higher uncertainty. The diamond (244-440) and the triangle (203-504) mark proplyds irradiated by $\theta^2$ Ori A. The dashed purple line represents a model prediction for disks with a radius of 40 au \citep{Storzer1999PhotodissociationOrion}.}
    \label{fig:G0-Mloss}
\end{figure}

To analyze how the estimated $\dot{M}_{\rm loss}$ of proplyds depends on the strength of the ionizing field, we plot the $\dot{M}_{\rm loss}$ values as a function of the incident UV radiation field in Fig.~\ref{fig:G0-Mloss}. This shows an increasing trend for $\dot{M}_{\rm loss}$ for the proplyds irradiated by $\theta^1$ Ori C, moving from 10$^6$ to 10$^7$ $G_0$. Proplyd 203-504 also follows the trend relatively well. The proplyds with uncertain measurements of the I-front radii have a systematically lower value of $\dot{M}_{\rm loss}$, but the uncertainty in the measurement does not allow further considerations.
We assessed the trend between $\dot{M}_{\rm loss}$ and $G_0$ with the statistical analysis described in the previous section. The Spearman correlation coefficient $r_s\simeq$~0.84, indicating a strong positive relation (Fig. \ref{fig-app:correlation}c), and p-value~$\simeq$~0.001, which indicates a very significant correlation.

\section{Discussion}\label{sect:discussion}

\subsection{Ratios of ionization front sizes}\label{sect:ratio_IF}

We have analyzed the MUSE IFU data of 12 proplyds to measure the I-front radii in four emission lines. The I-front radii as measured in different tracers reflect the ionization stratification in photoevaporating flow models \citep{Henney1998ModelingProplyds}, with the highest ionized flow spanning the furthest from the disk and closest to the UV source. Figure \ref{fig:O-ratios} underlines this ionization structure as R$_{IF}$[O\,\textsc{iii}] $>$ R$_{IF}$[O\,\textsc{i}], and
R$_{IF}$[O\,\textsc{iii}]/R$_{IF}$[O\,\textsc{i}] $>$ R$_{IF}$[O\,\textsc{ii}]/ R$_{IF}$[O\,\textsc{i}]. However, the statistical significance is limited by the small sample size.

We observe no dependency of the ratios on the impinging radiation, which indicates that the distribution of ionization front sizes of proplyds in the sample is self-similar: they have the same relative sizes at all levels of ionization, independently of the incident radiation. We provide a simple explanation of why this is expected, as follows.

Assuming equilibrium between ionizing flux from a massive star at distance $d$ and gas recombinations, another way to express R$_{IF}$, following the analytic treatment of flows by \citet{Johnstone1998PhotoevaporationNebula}, is:
 \begin{equation}
     r_{IF} = \left[\frac{\alpha r_0^4 n_0^2 }{3} \frac{4\pi d^2}{\dot{N}_i}\right] ^{1/3}
 \end{equation}
 where $\alpha$ is the recombination coefficient for the given ionized species, $r_0$ is the disk outer radius, $n_0$ is the density of neutral material at the disk surface, and $\dot{N}_i$ the ionizing photon count. When considering the ratio between I-front radii of two emission lines, the influence of distance becomes negligible, and the ratio depends on $\alpha$, and $\dot{N}_i$. Considering the same UV source, the ratio for [O\,\textsc{i}] and [O\,\textsc{ii}], for example, would simplify as 
 \begin{equation}
     \frac{r_{\rm IF,OII}}{r_{\rm  IF,OI}} = \frac{(\alpha_{\rm OII} \dot{N}_{\rm  OI} )^{1/3} }{(\alpha_{\rm OI} \dot{N}_{\rm OII} )^{1/3} }
 \end{equation}

In the case of a $\theta^1$\,Ori C like object, with $\dot{N}_{\rm OII}\sim10^{47}$, $\dot{N}_{\rm OI}\sim10^{49}$, $\alpha_{\rm OI}=3.25\times 10^{-13}$, $\alpha_{\rm OII}=1.99\times 10^{-12}$ (recombination coefficients from \citet{Draine2011PhysicsMedium}) the ratio is $\sim$9, so while it is expected to be insensitive to the distance of the proplyd from the UV source, the predicted magnitude of the ratio is too high from this simple analysis. The reason for this is that the above argument assumes a spherically diverging ($R^{-2}$) scaling of the density at all distances from the disk. In reality, at the hydrogen ionization front (which coincides closely with the singly ionized oxygen front) there is a sharp jump in temperature from tens/hundreds of Kelvin to $\sim10^4$\,K. That temperature jump is associated with a sharp increase in the flow velocity and a sharp decrease in the gas density (easily an order of magnitude). Furthermore, if material is being accumulated in a shell within the proplyd cusp then the density contrast between the proplyd and surrounding H\,\textsc{ii} region can be even higher, which we explore below with a basic numerical photoionization calculation. 

\subsection{Photoionization models of the ionization front structure}\label{subsect:ionization-model}
The structure of the H\,\textsc{ii} region around a $\theta^1$ Ori C type star is well understood to have an inner zone of more highly ionized species such as O\,\textsc{iii} and He\,\textsc{ii} surrounded by a zone where hydrogen is still ionized but, for example there is neutral helium and only singly ionized oxygen \citep[e.g.][]{ODell2001StructureNebula,ODell2017WhichNebula}. In the case of the proplyds, the transition from [O\,\textsc{iii}], to [O\,\textsc{ii}] to [O\,\textsc{i}] is very sharp, as shown in Sect.~\ref{sect:ratio_IF}, suggesting that the density is substantially higher than the ambient H\,\textsc{ii} region. 

Here we build a simple set of models to estimate the required density near the I-front in the proplyds. The purpose of these models is purely to determine the density of gas required to make the separation between the different oxygen emission features as small as observed. We are not aiming at producing a detailed model of the structure of the proplyds.

We ran 1D spherical photoionization equilibrium calculations with the \textsc{torus} Monte Carlo Radiative Transfer code \citep{Haworth2012RadiationField,Harries2019TheCode}. This included polychromatic radiation, the diffuse field. The species included are hydrogen, helium, oxygen, carbon, nitrogen, neon and sulfur. We set the dust abundance to a negligible value in these calculations.

Our model consists of a star with $T_{\rm eff}=$39,000\,K and radius $R_* = 10.6\,R_{\odot}$. The inner portion of the 1D spherical grid was then modeled to be representative of the ``normal'' H\,\textsc{ii} region, in which the gas density is $100\,$m$_{\rm H}$\,cm$^{-3}$. If the medium were all at that density, the H\,\textsc{ii} region would extend to around 3.3\,pc. We then imposed a density enhancement at 0.1\,pc from the UV source that is supposed to represent the denser medium associated with the proplyd cometary cusp. The increased density is then a free parameter of the models and we study the location of the oxygen ionization fronts. We note that it is not an issue if the separation between ionization fronts is not significantly smaller than the radius of curvature of the object in the 1D spherical model, since the required thickness for that to be the case would be inconsistent with the observed thickness for the real proplyds.

\begin{figure}
    \centering
    \includegraphics[width=\columnwidth]{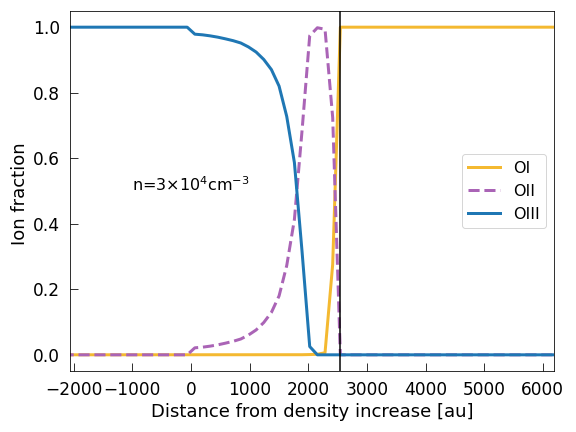}
    \includegraphics[width=\columnwidth]{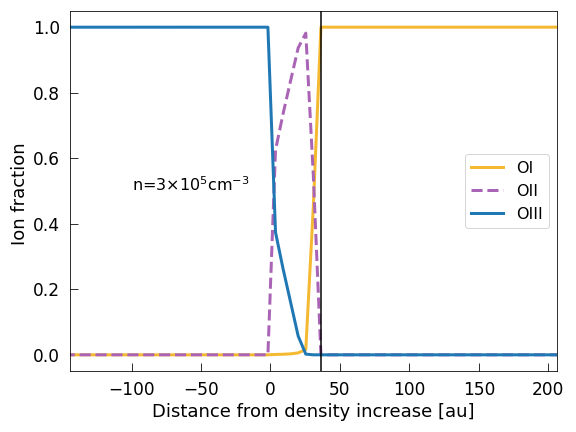}    
    \caption{Example profiles of the oxygen ionization state as a function of distance from the density enhancement in the H\,\textsc{ii} region (distance from the star of the proplyd cusp). The upper and lower panels are for densities of $3\times10^4$\,cm$^{-3}$ and  $3\times10^5$\,cm$^{-3}$ respectively.}
    \label{fig:Oexamples}
\end{figure}

Figure \ref{fig:Oexamples} shows the oxygen ionization fractions as a function of distance from the density enhancement (i.e., the distance outside of the normal H\,\textsc{ii} region and into the ``proplyd'') for different densities. At low densities the spacing between the different oxygen emitting zones is far larger than the extent of proplyds. At higher densities, at around a few $\times10^5$\,cm$^{-3}$, the model results in a spatial distribution of ionization states of oxygen consistent with what is typical for the observed proplyds.

\begin{figure}
    \centering
    \includegraphics[width=\columnwidth]{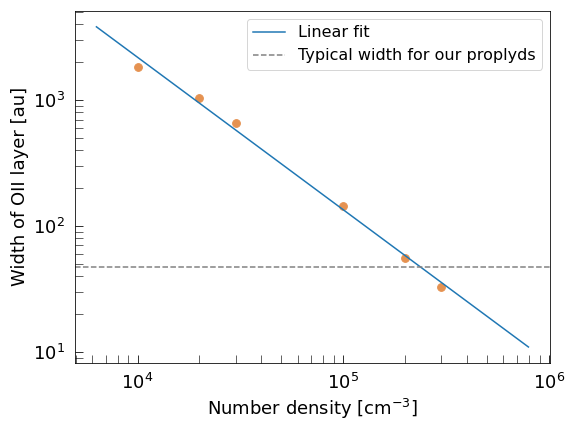}
    \caption{Distance between the O\,\textsc{iii}-O\textsc{ii} transition and the O\,\textsc{ii}-O\textsc{i} transition for a proplyd at a distance of 0.1\,pc from a $\theta^1$ Ori C like star, as a function of the density in the cometary cusp of the proplyd. For densities $\sim2-3\times10^5$\,cm$^{-3}$ the thickness of the O\textsc{ii} layer is consistent with our observations of ONC proplyds. }
    \label{fig:photoionModel}
\end{figure}

It is possible to broadly encapsulate the behavior here with the width of the O\,\textsc{ii} zone. Figure \ref{fig:photoionModel} shows the scaling of that width as a function of the density enhancement beyond 0.1\,pc. The scaling is approximately linear in log space, of the form
\begin{equation}
    \log_{10}\left(\frac{\Delta(\textrm{O}\,\textsc{ii})}{\textrm{au}}\right) = 8.18 - 1.21  \log_{10}\left(\frac{n}{\textrm{cm}^{-3}}\right) 
    \label{n_ROII_scaling}
\end{equation}
Although this was derived for a proplyd at 0.1\,pc from $\theta^1$ Ori C, this distance is quite representative for our sample, as the projected separation are lower limits of the real distances. Furthermore, if the proplyd is closer, the proplyd would have to be even denser than predicted by Eq. \ref{n_ROII_scaling} to achieve the same O\,\textsc{ii} width. The same would be the case if the H\,\textsc{ii} region density were lower. Overall these combine to make Eq. \ref{n_ROII_scaling} a lower limit, in such a way that the empirical scaling of Equation \ref{n_ROII_scaling} could be applied to give rough representative values of the density of cometary cusps across the ONC, which results in the typical values being $\sim2-3\times10^5$\,cm$^{-3}$. More detailed modeling is needed to precisely determine densities in proplyds using this approach, and we defer this analysis to future work. The densities predicted close to the I-front in our model are consistent with the electron density estimates inferred from radio continuum observations of ONC proplyds by \citet{Ballering2023IsolatingObservations}. 

\subsection{Mass-loss rate}\label{subsect-discussion:massloss}

Our estimates of the mass-loss rate employ the I-front radius, measured from the H$\alpha$ line emission, following Eq.~\ref{eq:RIFMdot}. The latter provides an upper limit on the mass-loss rate, since it neglects any possible extinction between the irradiating star and the proplyd. Indeed, clumps of dust between the ionizing source and the young stars can cause a total shielding of the ambient radiation field, as in the case of object 204-506, which does not show a cocoon of ionized material unlike the nearby 203-504 proplyd  \citep{Haworth2023TheBar}.

Figure \ref{fig:G0-Mloss} shows a decrease in the mass-loss rate with a smaller incident UV radiation field. This is in agreement with photodissociation models \citep{Storzer1999PhotodissociationOrion}, which consider that $\dot{M}_{\rm loss}$ is a slowly decreasing function of the UV field strength. The theoretical expectation is marked with the dashed line in Fig. \ref{fig:G0-Mloss}. This prediction reproduces the increase of $\dot{M}_{\rm loss}$ with incident radiation observed in our measurements, but at lower $\dot{M}_{\rm loss}$ values. The offset is likely due to several assumptions, such as dust extinction, the outflow velocity, and the disk radius. The latter has a value of $6 \times 10^{14}$ cm (40 au) in the plotted curve.

Proplyd 244-440 stands out as an outlier in Fig. \ref{fig:G0-Mloss} with $\dot{M}_{\rm loss}$ nearly an order and a half of magnitude larger than for the nearby proplyd 203-504. We discuss the properties of this object in the next subsection.

\subsubsection{The curious case of 244-440}\label{sect:discussion-244-440}
Proplyd 244-440 exhibits the widest cusp observed in the Trapezium. Past studies note that the giant proplyd is possibly irradiated by both $\theta^2$ Ori A and $\theta^1$ Ori C \citep{ODell2017WhichNebula}, evident from its multiple tails, although five times further from the latter (see Fig. \ref{fig:thesample}). The complex environment might play a role in the unusual shape of 244-440; \citet{Haworth2023TheBar} showed that the ONC is an environment where two YSOs can appear extremely different from each other regardless of their proximity. 

The disk diameter of 244-440 has been estimated to be 387~au from HST-WFPC2 images \citep{Vicente2005SizeCluster}. Using the higher spatial resolution observation of MUSE, the disk is roughly 300 au in diameter in the [O\,\textsc{i}]$\lambda$6300\AA\, line. This is a particularly large size for disks in the ONC, smaller than the giant disk 114-426 \citep[e.g.,][]{Miotello2012EVIDENCEDISK} but much larger than any other disk in this region, typically smaller than $\sim$60 au in radius \citep{Eisner2018ProtoplanetaryObservations}. 
For a disk size of that order, exposed to a high UV radiation field ($>10^5 G_0$), the models predict mass-loss rate $>10^{-6}$\,M$_\odot$\,yr$^{-1}$ \citep{Haworth2023scpfried/scpDiscs}, consistent with our estimation. 

Despite standing out from the sample with its large I-front radius and high mass loss rate, 244-440 follows the same behavior as the rest of the objects in Fig. \ref{fig:O-ratios}, with the ratios of its I-front radius being similar to the rest of the sample. This confirms a similar effect of the irradiation from the massive star(s) on the relative sizes of the I-fronts, as expected by the fact that they depend mainly on physical constants and, to a lesser extent, on the densities in the systems (see Sect.~\ref{sect:ratio_IF}-\ref{subsect:ionization-model}). 

Given the rapid mass-loss rate, 244-440 is a unique object to be observed before its dispersal.
Future observations with ALMA would clarify the disk properties and give insight into the evolution of the proplyd.\\

\begin{figure}[]
    \centering
\includegraphics[width=\columnwidth]{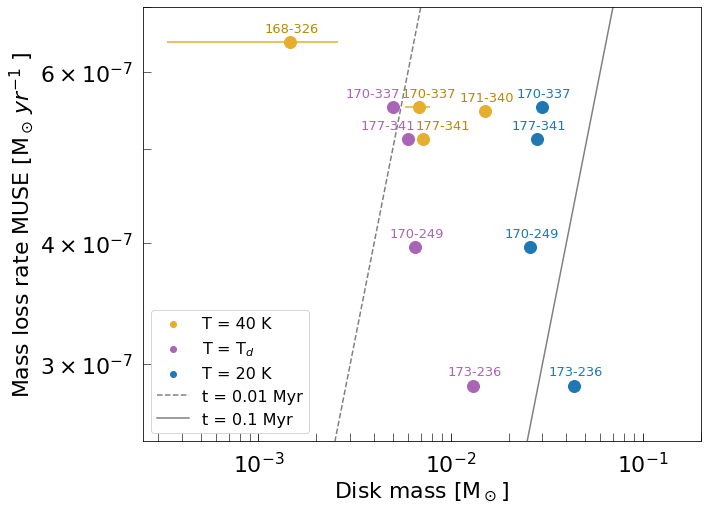}
    \caption{Mass-loss rates measured from the ionization front radius as a function of disk mass. The disk masses found by \citet{Mann2014ALMAPROPLYDS} partly overlap with those reported by \citet{Ballering2023IsolatingObservations}. Each colour represents a different assumed dust temperature: T=20 K, T=40 K, and T$_d$=62--108 K, and diagonal lines mark lifetimes.}
    \label{fig:Mdisk-Mloss}
\end{figure}

\subsubsection{Dependence of mass loss rate on stellar mass}
We have determined the stellar parameters for 10 targets in our sample. One half of the sample is classified as M spectral type (M0 and M5), and the other half is classified as mid-late K spectral type. The stellar masses range from 0.2 to $\sim$1 M$_\odot$. 

It is known that, in such a stellar mass range, several disk, accretion, and wind properties have a steep scaling with stellar mass \citep[e.g.,][]{Manara2023DemographicsFormation}. Although this sample is small, a correlation between $\dot{M}_{\rm loss}$ and stellar mass would be expected if the mass loss rate was driven by other processes than the irradiation from the ionizing stars, or if the stellar mass was a significant factor in setting the strength of the photoevaporating wind.

We explore the relation between $\dot{M}_{\rm loss}$ and stellar mass and find no correlation or dependence between the two parameters, which implies that mass loss is driven by the external UV source and depends on other properties, such as disk radii, but not (significantly) on the stellar mass. On the other hand, it is known that, in low-mass star-forming regions, the disk radii typically scale with stellar mass \citep[e.g.,][]{Hendler2020TheStars}. This is not the case in the ONC. Indeed, \citet{Vicente2005SizeCluster} presented the disk diameter as a function of the SpT in a sample of 52 sources, and found no obvious correlation between the two.

Recently, \citet{Mauco2023TestingMeasurements} observed that the depletion of disk masses in the vicinity of the massive star $\sigma$-Ori is more pronounced for disks around stars with stellar masses higher than $\sim 0.4\,M_\odot$. Our sample is too small to firmly confirm or dispute this finding, but we do not find any significant difference in the measured $\dot{M}_{\rm loss}$ for the disks around stars more massive than 0.4 $M_\odot$. The reason could be a combination of the location and impinging radiation onto the disks, or the lack of a dependence on stellar mass. Future studies measuring both $\dot{M}_{\rm loss}$ and stellar masses should further explore this possibility.

\subsection{Proplyd lifetime}
The external photoevaporation process erodes material from the outer part of protoplanetary disks at a rate calculated here for our sample as $\dot{M}_{\rm loss}$. Here we investigate how long the disks we have observed can appear to survive under the effect of this process.
In Fig. \ref{fig:Mdisk-Mloss}, we show $\dot{M}_{\rm loss}$ as a function of the disk masses measured with ALMA from continuum emission for seven proplyds from \citet{Mann2014ALMAPROPLYDS} (reported as 100 $\times$ dust mass) and \citet{Ballering2023IsolatingObservations}. The disk masses in the literature have been estimated by using different dust temperatures: \citet{Mann2014ALMAPROPLYDS} assumed 40 K, while \citet{Ballering2023IsolatingObservations} computed the disk masses at both 20 K, and at temperatures that take external heating \citep{Haworth2021WarmStars} by \thetaonec{} into account for each proplyd in their sample. 
The values of disk masses for our targets are lower than typical disk populations in nearby low-mass star-forming regions. Indeed, the relation between disk masses and the projected separation from \thetaonec{} have been studied by for example, \citet{Mann2010ACLUSTER}, who found that the disks located within 0.3 pc from the massive star have lower masses.
The low disk masses close to the UV source is also consistent with studies in different regions with lower but not negligible UV field, such as $\sigma$ Orionis \citep{Ansdell2017AnCluster,Mauco2023TestingMeasurements} and NGC 2024 \citep{Mann2015PROTOPLANETARYCLUSTER,vanTerwisga2020ProtoplanetaryPopulations}. Interestingly, also in the latter region proplyds were observed with HST \citep{Robberto2023A2024}. The $\dot{M}_{\rm loss}$ values are within the same order of $10^{-7}$\,M$_\odot$\,yr$^{-1}$ for the available disk masses ($10^{-4} M_\odot - 10^{-1} M_\odot$). 

The ratio between the disk mass and $\dot{M}_{\rm loss}$ gives an estimate of the expected lifetime of these systems, assuming that a constant $\dot{M}_{\rm loss}$ will affect them during their lifetime. Since other processes, such as accretion onto the star or internal winds, are predicted to remove mass from the disk at a rate that is orders of magnitudes lower \citep[e.g.,][]{Alcala2017X-shooterLupus,Manara2017X-ShooterStars,Nisini2018ConnectionStars}, this ratio is the most constraining for the disk lifetime. 
Observational studies report mass loss ($10^{-7}$--$10^{-6}$ M$_\odot$\,yr$^{-1}$) similar to our estimates, and lifetimes $<$ 1 Myr \citep[e.g.,][]{Storzer1999PhotodissociationOrion, Henney2002Mass2,Ballering2023IsolatingObservations} for the proplyds. Here we have a larger sample of proplyds with both measured $\dot{M}_{\rm loss}$ and disk mass, and we derive photoevaporative timescale $t=M_{\rm disk}$/$\dot{M}_{\rm loss}$ in the range from 2.4 to 130 kyr (Fig.~\ref{fig:Mdisk-Mloss}). These values are very short compared to the age of the region, making it peculiar that these objects are observed just before their final dispersal. 
Given these derived values, our results are in agreement with the literature values, conforming to this "proplyd lifetime problem".

Multiple works have explored solutions to the lifetime problem.
\citet{Storzer1999PhotodissociationOrion} suggested that disks can survive much longer if they have radial orbits, and experience the highest mass-loss rate when they are closer to the ionizing stars, when they fly by in their vicinity. However, the dynamical models of the ONC were not found to be consistent with the suggestion \citep{Scally2001DestructionCluster}.

The theoretical study by \citet{Winter2019AProblem} explored the survival of disks in the ONC by implementing a one-dimensional viscous evolution model, and highlighted that the mass loss rates inferred from observations of proplyds in the ONC are preferentially high, representing the most extreme values. Their model predicts $\dot{M}_{\rm loss}$ to be a factor of 2--3 lower for the majority of proplyds in their sample.
The study also shows that part of the solution involves accounting for the age spread present in the stellar population of the ONC \citep[e.g.,][]{Hillenbrand1997OnCluster,Palla1999StarCluster,DaRio2010Asup/sup,Jeffries2011NoCluster,Beccari2017ACities}. In this scenario, the strongly irradiated stars are also the youngest and therefore have the greatest mass reservoir remaining. This represents a natural mechanism by which disks in the ONC can survive until the present day. 
\citet{Wilhelm2023RadiationRegions} focused on modeling the radiation shielding by gas in massive star-forming regions, and showed that the disk lifetime could be extended by an order of magnitude by such shielding, making the lifetime problem milder. 
In the context of recent observational analysis, \citet{Ballering2023IsolatingObservations} suggested that disk masses are underestimated as a result of the ALMA emission being optically thick, which could play a role in solving the proplyd lifetime problem.
Therefore, there are multiple factors to take into account when estimating the lifetimes of proplyds.

\section{Conclusions}
We have presented the first spatially and spectrally resolved observations of 12 proplyds in the Orion Nebula Cluster using MUSE NFM. We have characterized their morphology in different emission lines commonly associated with ionized gas, measured the size of the ionization front in different ionization tracers, and measured their stellar properties. This has allowed us to estimate physical parameters and draw comparisons with photoevaporation models.
Our main conclusions are as follows:

\begin{itemize}
   \item The ionization front radius traced by different emission lines (\Ha, [O\,\textsc{i}], [O\,\textsc{ii}], and [O\,\textsc{iii}]) increases with decreasing strength of the FUV radiation field, as expected by photoevaporation models. 
    
    \item The ratios of the ionization front radii for the emission lines of H${\alpha}$, [O\,\textsc{i}] 6300\AA, [O\,\textsc{ii}] 7330\AA,  and [O\,\textsc{iii}] 5007\AA, shows the expected ionization stratification: the most ionized part of the flow is located closest to the UV source.
    
    \item The ratios of ionization fronts at the different ionization states explored in this work scale in the same way for all proplyds in our sample regardless of the incident radiation. This indicates that the ratio of ionization front sizes at different ionization states of proplyds is self-similar.

    \item Mass-loss rates, estimated based on the H${\alpha}$ emission, are of the order of 1.07--94.5 $\times$ 10$^{-7}$ M$_\odot\,{\rm yr}^{-1}$. Higher $\dot{M}_{\rm loss}$ is observed at higher FUV field strengths, in line with general expectations from models. No dependence is observed with the stellar mass.
    
    \item The density of cometary cusps across the ONC should have typical values $\sim$ 2--3$\times$10$^5$ cm$^{-3}$ to explain the observed spatial distribution of ionization states of oxygen in proplyds. 

    \item Observationally, the proplyd lifetime problem still holds even after new measurements of disk masses and mass loss rates have been performed.
    
\end{itemize}

This work sets the stage for identifying firm signatures associated with external photoevaporation, which would open the possibility of confirming whether disks located in more massive and distant star-forming regions undergo the effect. Moreover, the characterization of external photoevaporation is fundamental to constraining planet formation models.

Future work will involve analyzing line ratios as the tracers of electron density and temperature. Determining the physical properties of the proplyds will allow for a more precise measurement of the mass-loss rate. Additionally, the identification of new emission lines will be explored.

\begin{acknowledgements}
We thank the anonymous referee for the insightful review that improved our study.
MLA, KM, and CFM acknowledge funding from the European Union (ERC, WANDA, 101039452).
SF is funded by the European Union (ERC, UNVEIL, 101076613) and by the grant PRIN-MUR 2022YP5ACE. GPR is funded by the Fondazione Cariplo, grant no. 2022-1217, and the European Union (ERC, DiscEvol, 101039651).
TJH acknowledges funding from a Royal Society Dorothy Hodgkin Fellowship and UKRI guaranteed funding for a Horizon Europe 
ERC consolidator grant (EP/Y024710/1). Views and opinions expressed are however those of the author(s) only and do not necessarily reflect those of the European Union or the European Research Council. Neither the European Union nor the granting authority can be held responsible for them.
\end{acknowledgements}

% APPENDIX:
% show the plots of the I-front measurement for all proplyds 

% \clearpage
\bibliographystyle{aa}
\bibliography{references}

\begin{appendix}

\section{MUSE NFM images of ONC proplyds}

\begin{figure*}
    \centering
    \includegraphics[width=0.95\textwidth]{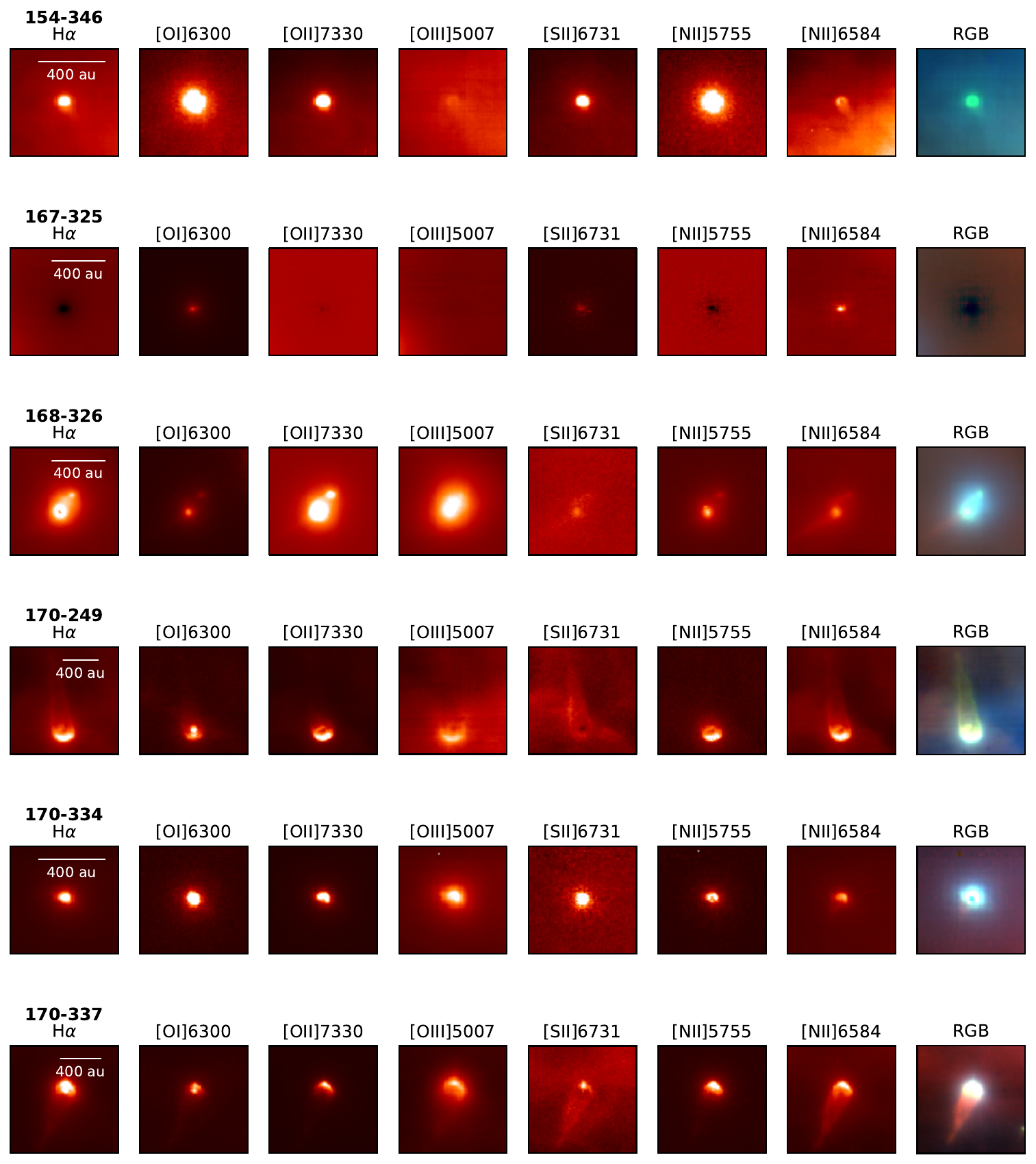}
    \caption{MUSE NFM observations of proplyds in seven emission lines. The last column is an RGB image combining [N\,\textsc{ii}] 6548\AA, \Ha, and [O\,\textsc{iii}] 5007\AA\ emission lines.}
    \label{fig-app:collage-1}
\end{figure*}

\begin{figure*}
    \centering
    \includegraphics[width=0.95\textwidth]{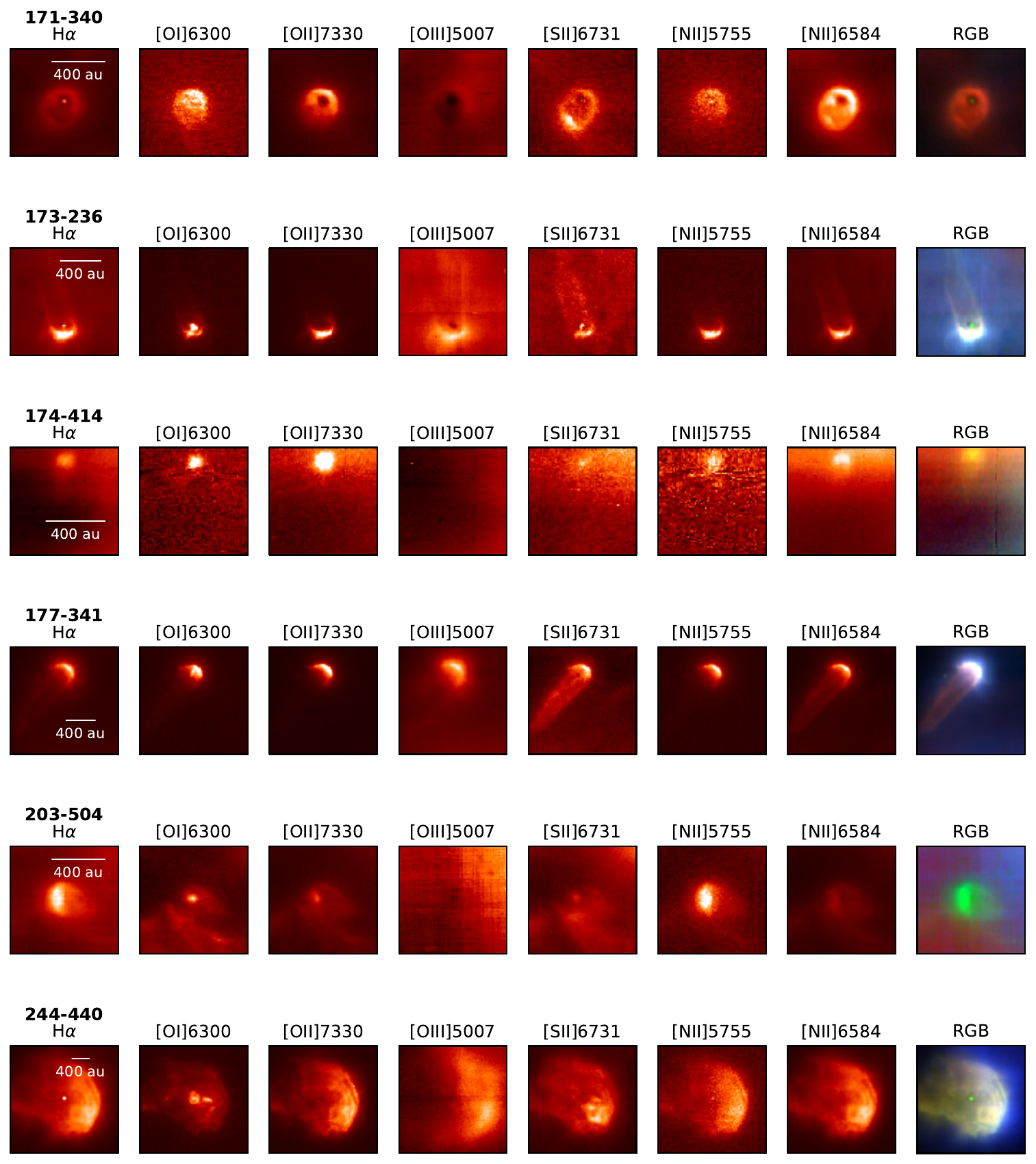}
    \caption{MUSE NFM observations of proplyds in seven emission lines. The last column is an RGB image combining [N\,\textsc{ii}] 6548\AA, \Ha, and [O\,\textsc{iii}] 5007\AA\ emission lines.}
    \label{fig-app:collage-2}
\end{figure*}

Figures \ref{fig-app:collage-1} and \ref{fig-app:collage-2} show each proplyd in a collage of the following emission lines: \Ha, [O\,\textsc{i}] 6300\AA,  [O\,\textsc{ii}] 7330\AA,  [O\,\textsc{iii}]~5007\AA, [S\,\textsc{ii}] 6731\AA, [N\,\textsc{ii}] 5755\AA, and [N\,\textsc{ii}] 6584\AA. These lines are selected to show the various components of the systems.

\section{Statistical test}
To asses the statistical significance of the relations in our small sample, we carry out two analyses for the following: ionization front radius ($R_{IF}$) and projected separation (d); R$_{IF}$ with UV radiation ($G_0$); and mass-loss rate ($\dot{M}_{\rm loss}$) with $G_0$.

\begin{figure*}
    \centering
    \includegraphics[width=0.99\textwidth]{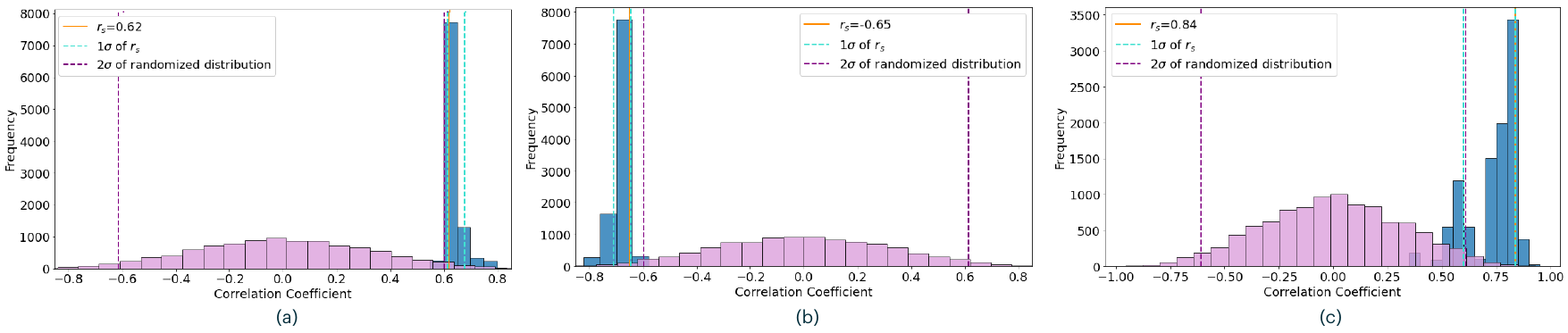}
    \caption{Correlation coefficient analysis for: (a) the ionization front radii and projected separation relation; (b) the ionization front radii and UV radiation relation; (c) mass-loss rate and UV radiation relation.
    The violet histogram corresponds to the bootstrapped randomized x-axis values; the dashed purple lines mark the 2$\sigma$ ranges of the realizations.
The blue histogram results from bootstrapping the y-axis values within the respective error bars, and the 1$\sigma$ ranges of the bootstrap are shown with cyan line.
We show that Spearman correlation coefficient for the measured data (orange line), which is outside 2$\sigma$ of the randomized distribution.}
    \label{fig-app:correlation}
\end{figure*}

Firstly, we test the null-hypothesis that there is no relation between the independent variable on the x-axis and the dependent variable on the y-axis. We use a bootstrap sampling method with 10$^4$ iterations, where we randomize the x values and compute the Spearman correlation coefficient. Secondly, we assess the uncertainty of the correlation coefficient by bootstrapping with 10$^4$ iterations while varying the y-axis values within their respective uncertainties taken as the standard deviation of a Gaussian distribution of random points. We adopt a truncated Gaussian distribution to model the uncertainty for the lower limits values in the $R_{IF}$, and thus also $\dot{M}_{\rm loss}$, measurements. In those cases, the bootstrapping is selected within this truncated Gaussian distribution with a standard deviation of 20\% of the lower limit value. Both of these tests are then compared with the actual Spearman correlation coefficient $r_s$. The results are presented in Fig. \ref{fig-app:correlation}.

\section{Best fit of the stellar spectra}
\begin{figure*}[]
    \centering
\includegraphics[width=\columnwidth]{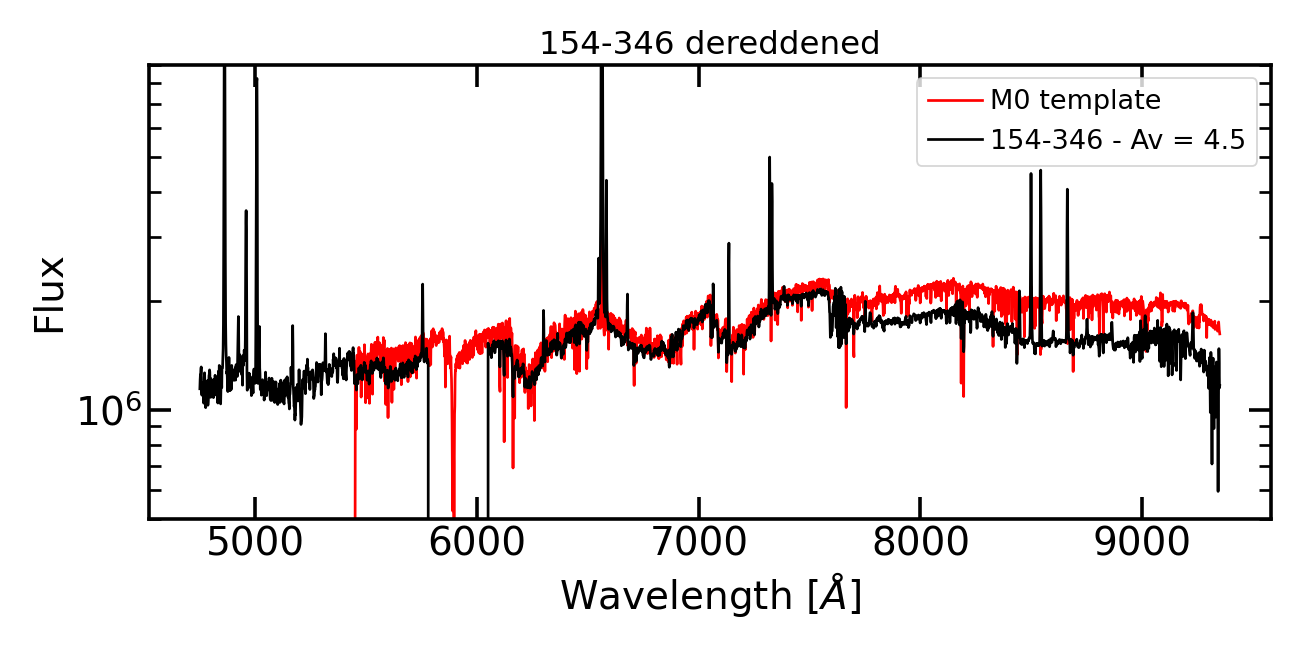}
\includegraphics[width=\columnwidth]{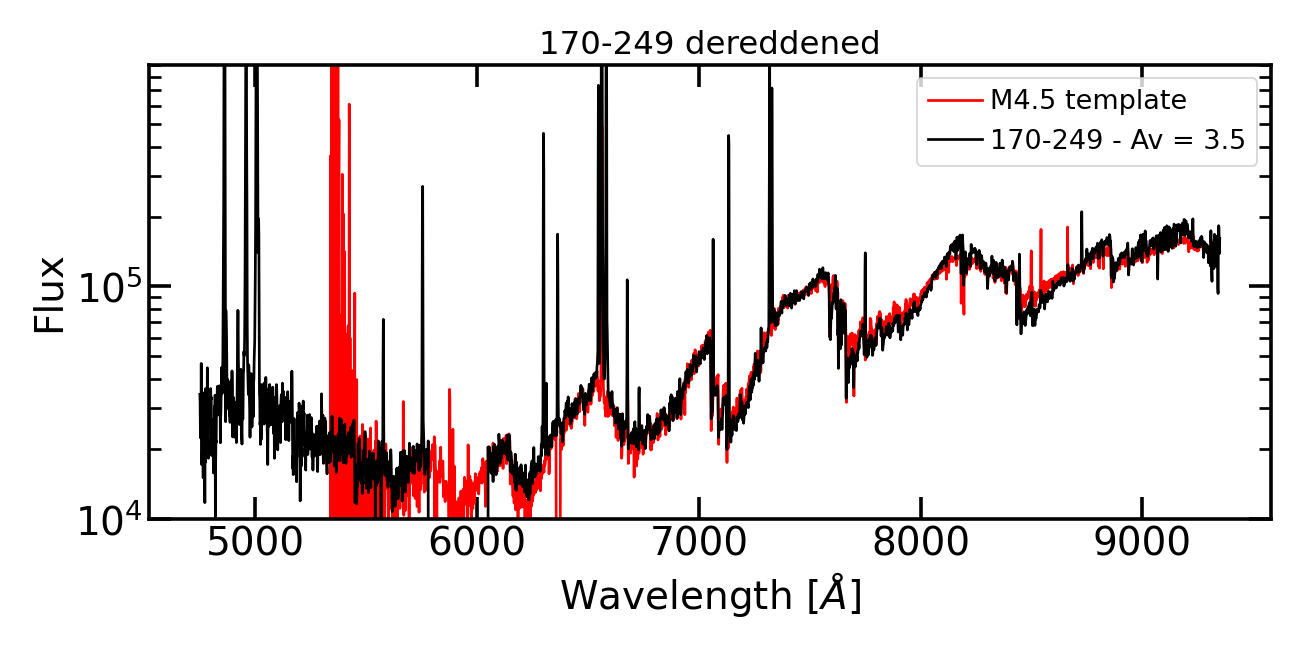}
\includegraphics[width=\columnwidth]{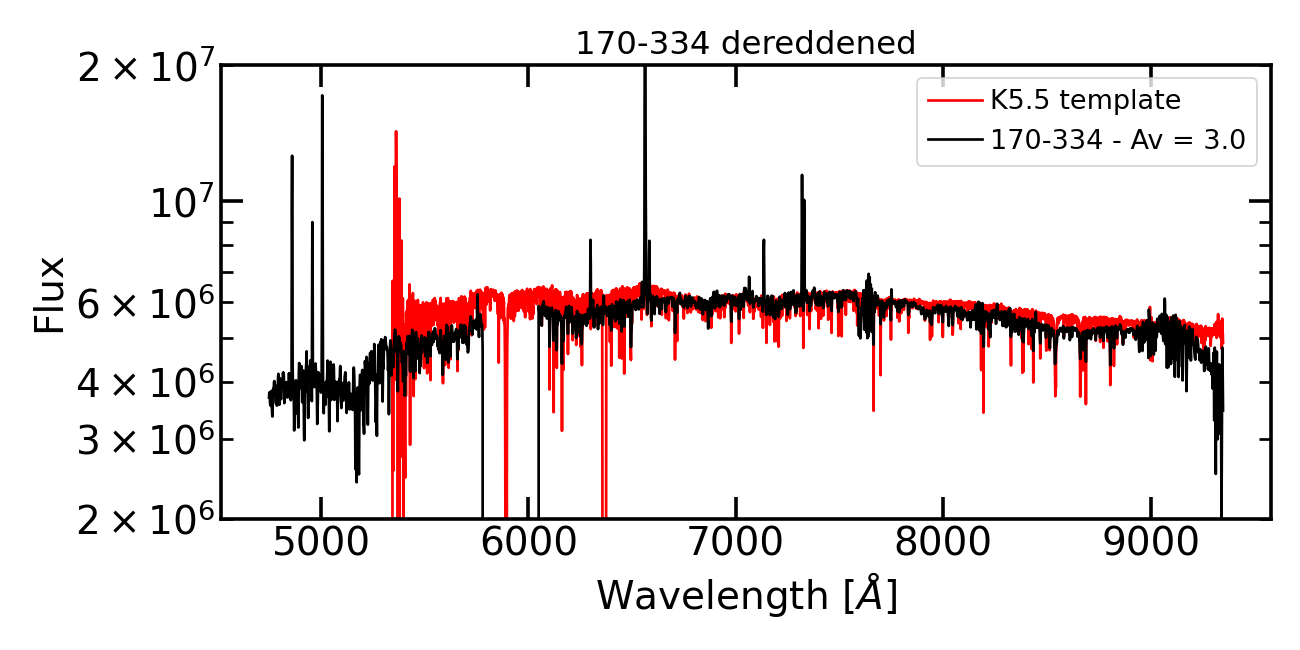}    
\includegraphics[width=\columnwidth]{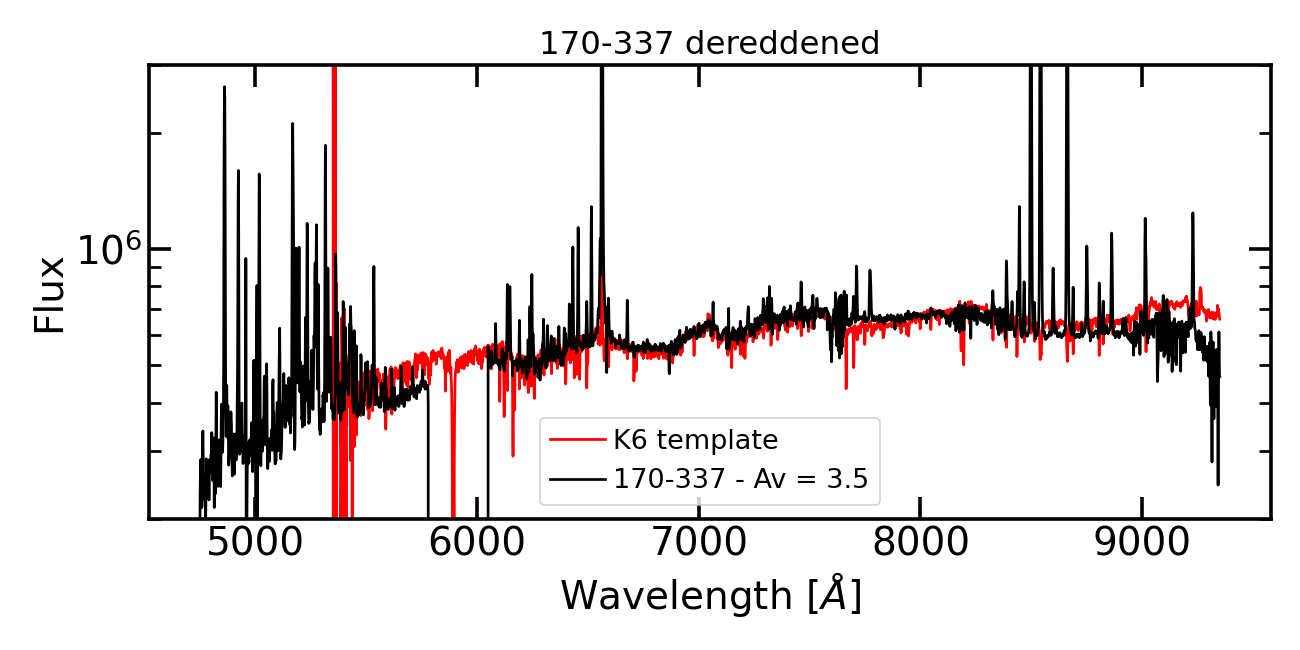}    
\includegraphics[width=\columnwidth]{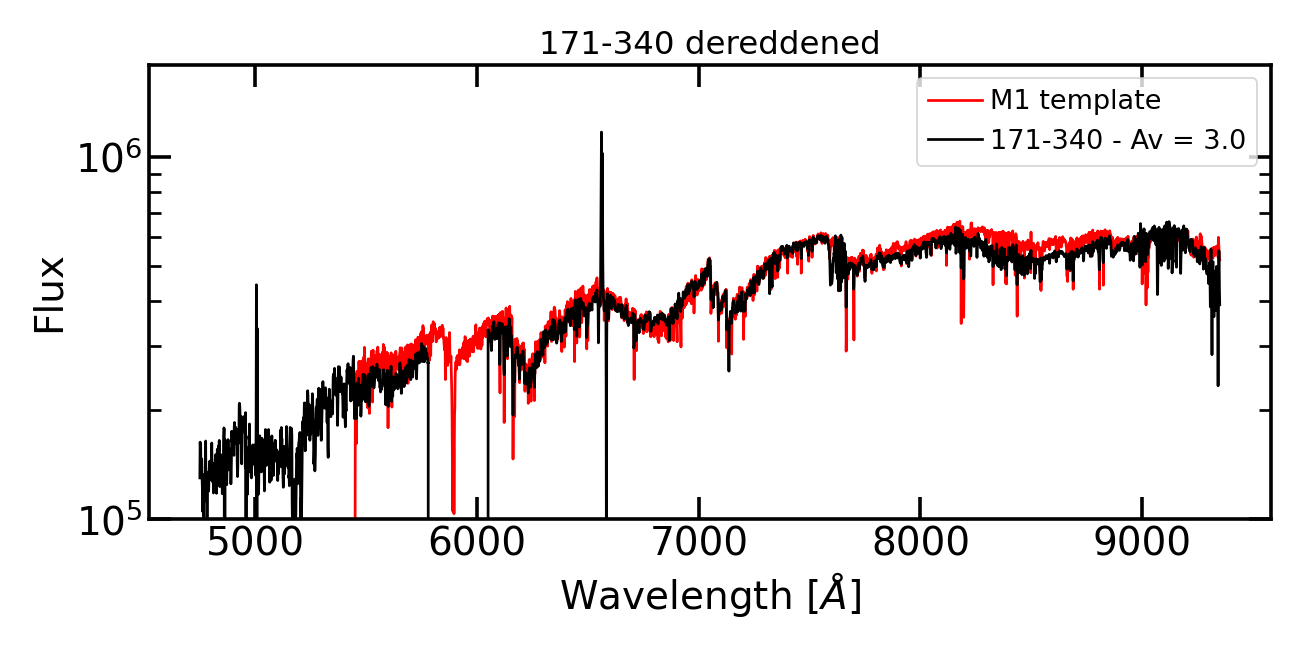}    
\includegraphics[width=\columnwidth]{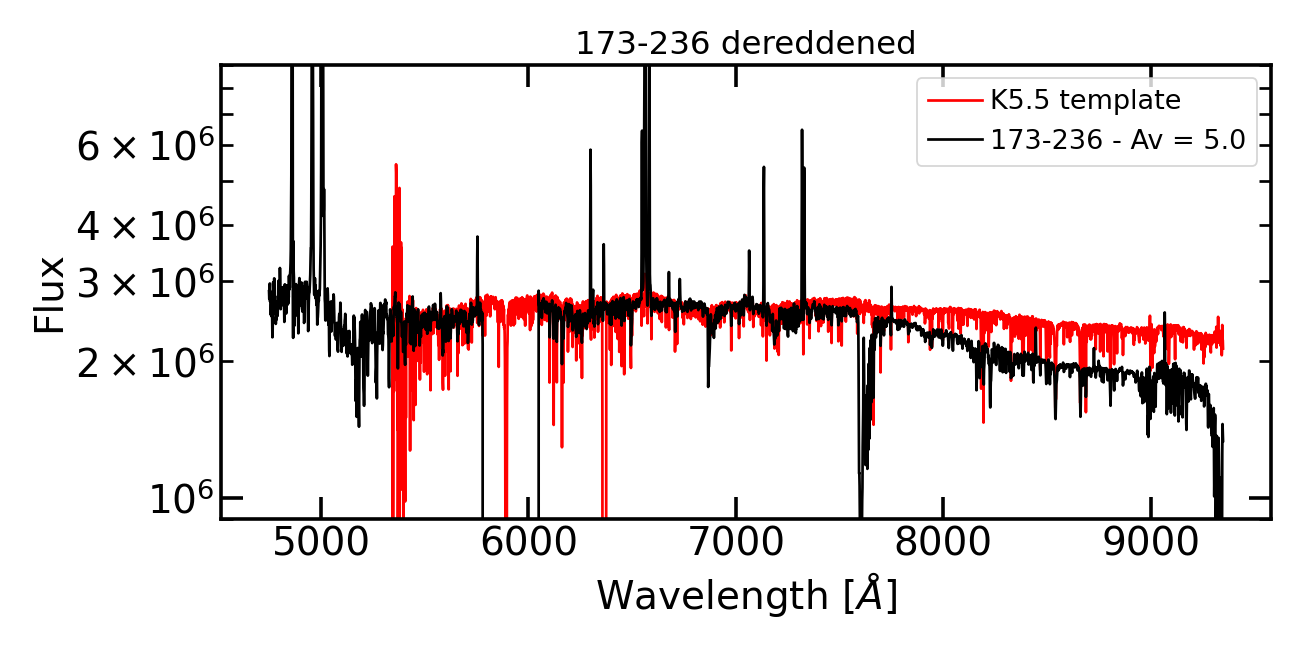}    
\includegraphics[width=\columnwidth]{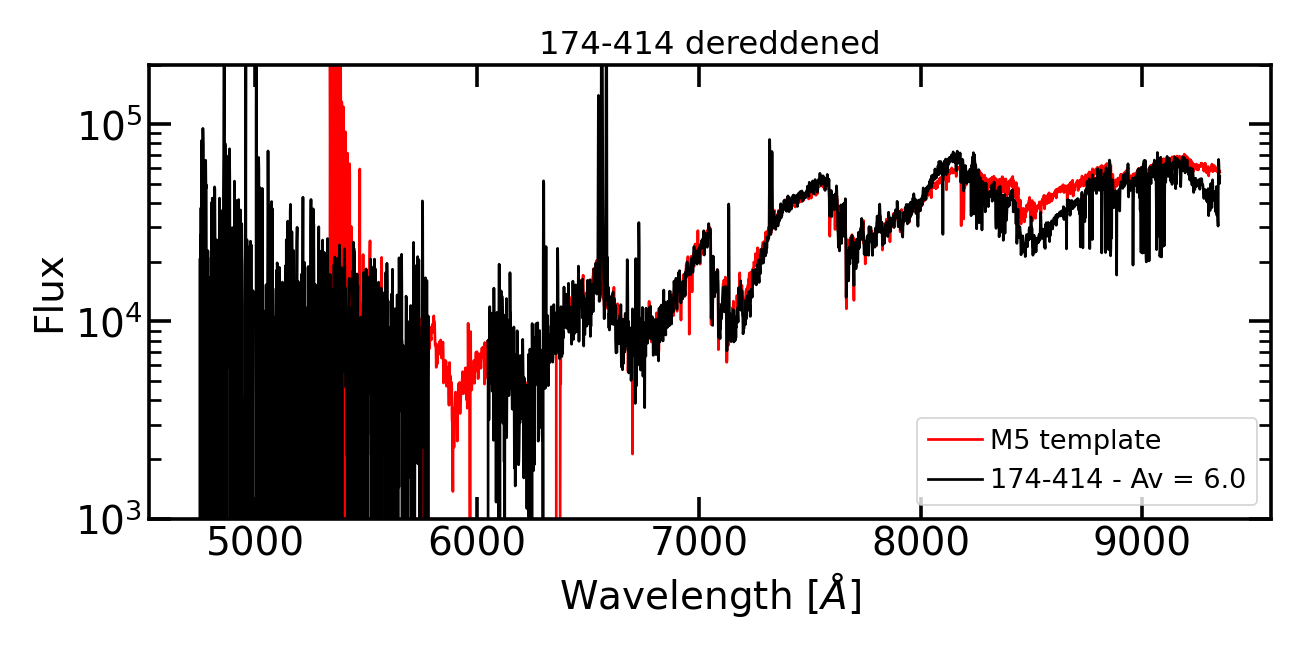}    
\includegraphics[width=\columnwidth]{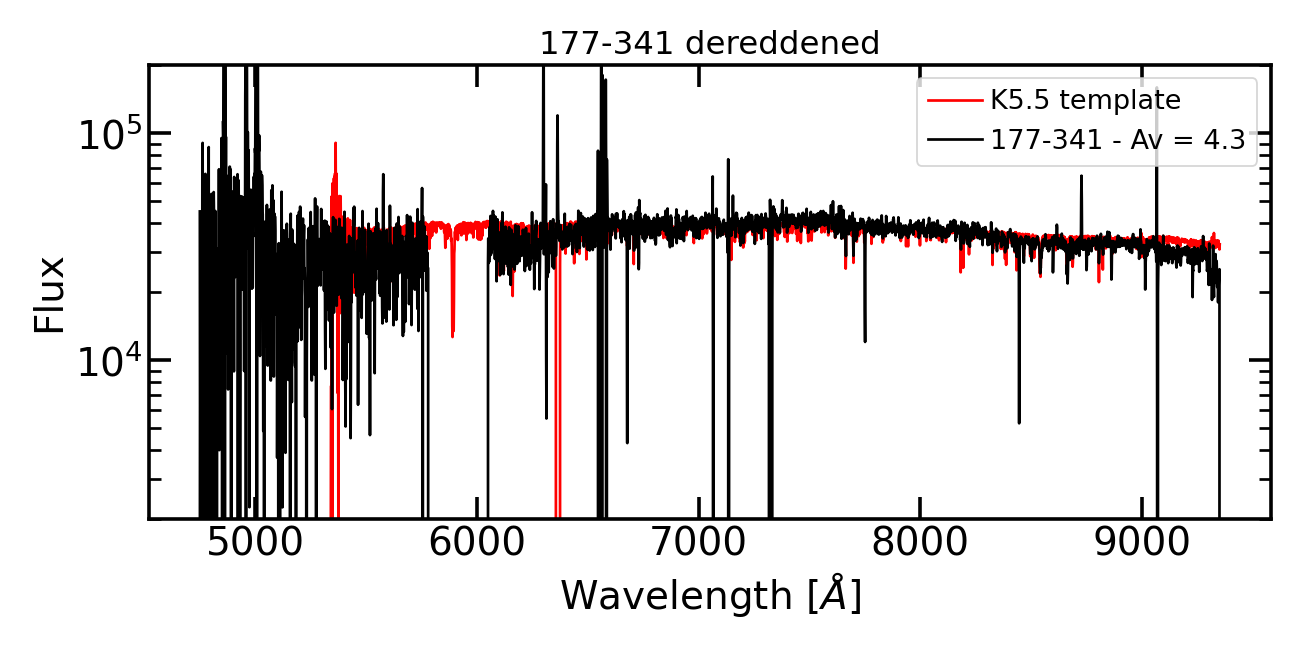}    
\includegraphics[width=\columnwidth]{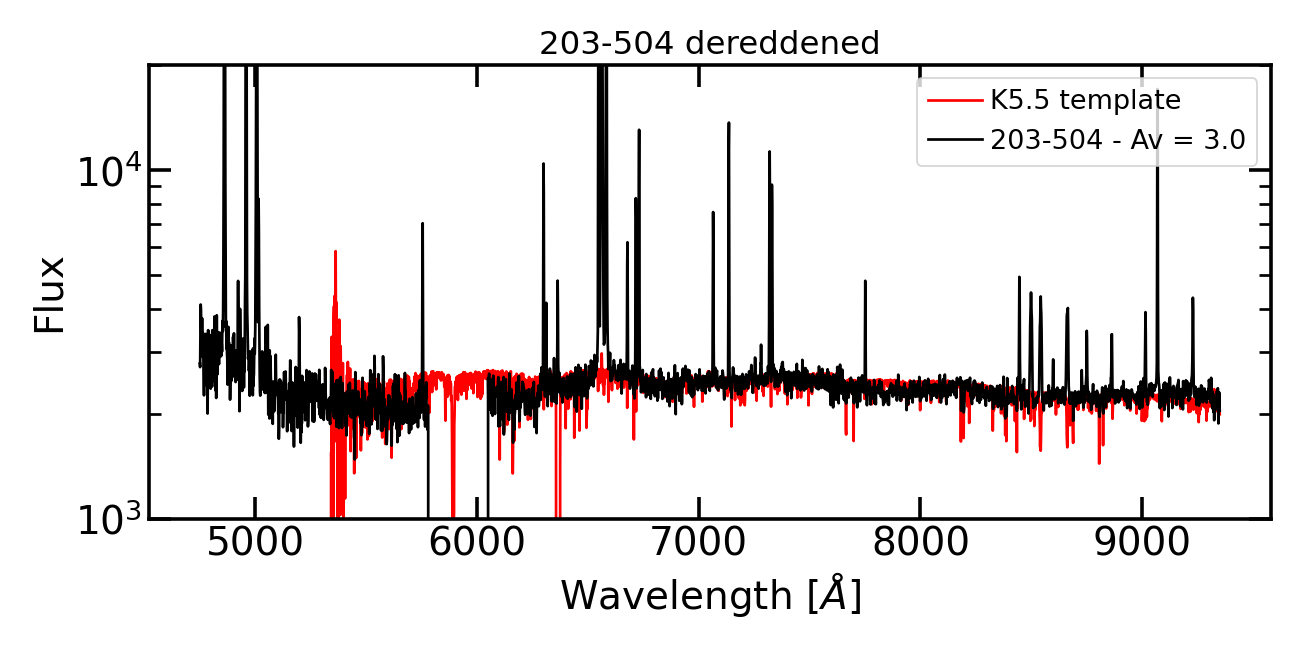}    
\includegraphics[width=\columnwidth]{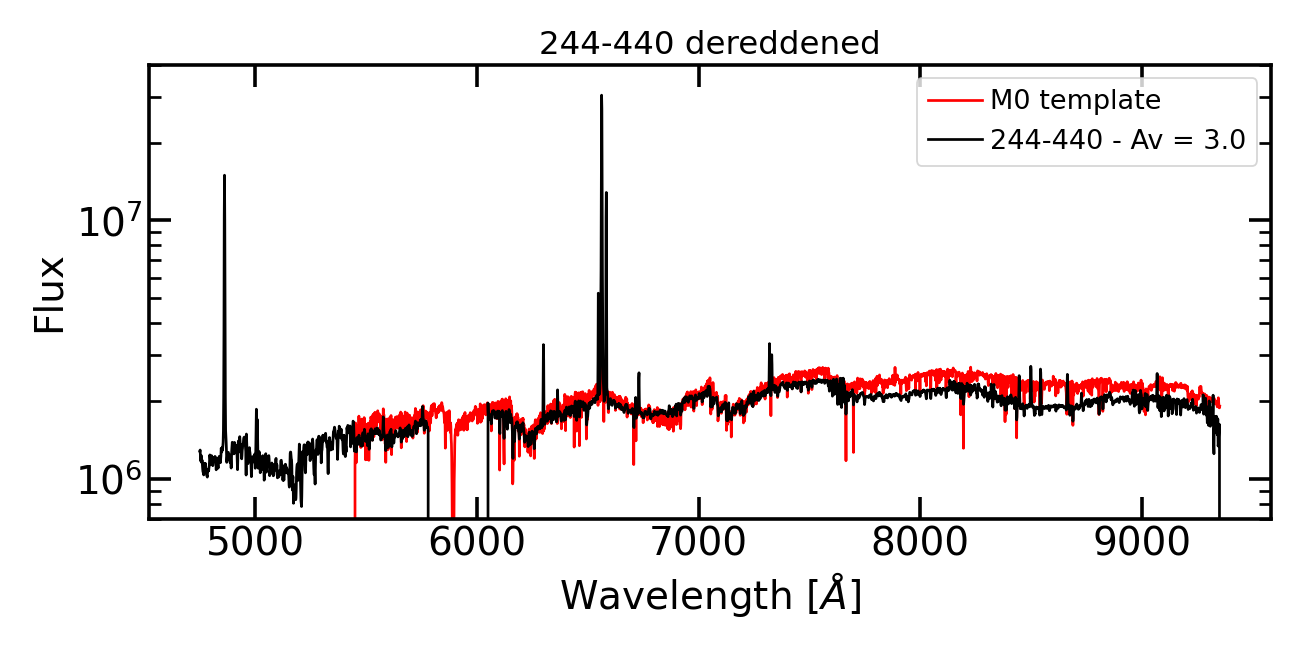}    
\caption{Best fit of the stellar spectra of the proplyds studied in this work}
    \label{fig:best_fit_spt-1}
\end{figure*}

As described in Sect.~\ref{sect::star_properties}, the extracted stellar spectra for the proplyds are compared with templates of young stellar objects to determine their spectral type and extinction. The best fit of the spectra are presented in Fig.~\ref{fig:best_fit_spt-1}. 

\newpage

\section{Observations}
The details concerning the MUSE NFM observations are given in Table \ref{Table:Obs-Log}.
% \begin{table*}

\begin{sidewaystable}
\caption{Information about the observations taken with MUSE NFM}
\begin{tabular}{c c ccccccccccc}
% \begin{tabular}{c c ccccccccccc}
\toprule
\toprule
Proplyd Name & Observation date  &  Exp. Time &  Seeing &  NEXP & Airmass Start &  Airmass End &  Seeing Start & Seeing End &  Tau0 &  Pr. Id \\
(1) & (2) & (3) & (4) & (5) & (6) & (7) & (8) & (9) & (10) & (11) \\
 \midrule
    142-301 & 2021-01-19 T04:37:43.619 &               360 &     0.84 &         3 &       1.262 &     1.288 &        0.89 &       0.8 &  0.005 &  A \\
    159-350 & 2021-01-21 T01:40:53.355 &               729 &     0.81 &         3 &       1.071 &     1.067 &         0.5 &      0.52 &  0.007 &  A \\
    167-231 & 2021-01-22 T02:17:52.677 &               340 &     0.62 &         3 &        1.06 &     1.061 &        0.64 &      1.08 &  0.005 &  A \\
    168-326 & 2019-10-12 T07:53:02.561 &               400 &     1.25 &         3 &       1.098 &     1.089 &        1.13 &      0.74 & 0.0027 & B \\
    170-249 & 2021-02-20 T02:00:13.622 &               360 &     0.71 &         3 &        1.17 &     1.188 &        0.52 &      0.54 &  0.007 &  A \\
    170-337 & 2021-01-21 T02:01:15.202 &               300 &     0.74 &         3 &       1.061 &      1.06 &        0.41 &      0.79 & 0.0065 &  A \\
    173-236 & 2021-01-25 T03:54:24.967 &               300 &     0.84 &         2 &       1.201 &     1.218 &         0.6 &       0.6 & 0.0055 &  A \\
    177-341W & 2019-10-23 T07:23:13.289 &               380 &     0.82 &         3 &       1.083 &     1.076 &         0.9 &      1.01 & 0.0076 & B \\
    180-331 & 2021-02-06 T02:44:54.948 &               360 &     0.87 &         3 &       1.147 &     1.163 &        0.67 &      0.86 & 0.0084 &  A \\
    203-504 & 2022-11-20 T04:45:50.432 &               660 &     0.91 &         4 &       1.154 &     1.131 &        0.55 &      0.57 & 0.0065 &  C \\
    244-440 & 2019-10-23 T08:14:06.908 &               360 &     0.99 &         3 &       1.059 &      1.06 &         1.2 &      0.96 & 0.0063 & B \\
        HST-10 & 2021-03-25 T01:03:00.499 &               360 &     0.78 &         1 &       1.446 &     1.483 &        0.65 &      0.59 & 0.0043 &  A \\
\hline 
\end{tabular}
\tablefoot{Columns: (1) object,
(2) date of observations,
(3) exposure time per exposure,
(4) airmass corrected seeing,
(5) number of exposure,
(6) airmass at start,
(7) airmass at end of obs,
(8) seeing at start,
(9) seeing at end of obs,
(10) coherence time,
(11) Proposal ID. A: 106.218X.001. B: 104.C-0963(A). C: 110.259E.001.}\label{Table:Obs-Log}
\end{sidewaystable}

\end{appendix}

\end{document}